\documentclass{article}
\usepackage{graphicx} 
\usepackage{dsfont}
\usepackage{amsmath,mathrsfs}
\usepackage{geometry}[30mm]
\usepackage{jheppub}
\usepackage{bm}
\usepackage{tikz}
\usetikzlibrary{arrows.meta, positioning}

\title{Carroll and AdS/CFT in higher dimensions}
\author{}
\date{}

\newcommand{\eal}[1]{\be \begin{aligned} #1 \end{aligned}\end{equation}} 
\newcommand{\eqn}[1]{\be #1 \end{equation}} 
\newcommand{\eqa}[1]{\bea  #1\end{eqnarray}}

\newcommand{\mbo}{\mathbb{O}}

\newcommand{\bz}{\bar{z}}

\newcommand{\scri}{\mathscr{I}}

\newcommand{\an}[1]{\left\langle#1\right\rangle}
\newcommand{\mo}{\mathcal{O}}
\newcommand{\mc}{\mathcal{C}}
\newcommand{\mA}{\mathcal{A}}

\newcommand{\zb}{\bar{z}}
\newcommand{\Zb}{\bar{Z}}

\newcommand{\hypf}{{}_2F_{1}}

\newcommand{\D}{\Delta}
\newcommand{\e}{\epsilon}
\newcommand{\om}{\omega}

\long\def\new#1\endnew{{\bf #1}}		
\long\def\del#1\enddel{}

\def\del{\partial}

\numberwithin{equation}{section} 

\numberwithin{equation}{section} 
\usepackage{hyperref}
\usepackage{orcidlink}

\begin{document}

\begin{titlepage}

  \thispagestyle{empty}

  \begin{center}


{\LARGE\textbf{On Carrollian and Celestial Correlators}}
\vskip0.2cm
{\LARGE\textbf{in General Dimensions}}

\vskip 1cm
Harshal Kulkarni\footnote{\fontsize{8pt}{10pt}\selectfont\ \href{mailto:harshal.kulkarni@maths.ox.ac.uk}
{harshal.kulkarni@maths.ox.ac.uk}}, 
Romain Ruzziconi\footnote{\fontsize{8pt}{10pt}\selectfont\ \href{mailto:Romain.Ruzziconi@maths.ox.ac.uk}{romain.ruzziconi@maths.ox.ac.uk}}, 
Akshay Yelleshpur Srikant\footnote{\fontsize{8pt}{10pt}\selectfont \ \href{mailto:Akshay.YelleshpurSrikant@maths.ox.ac.uk}{yelleshpursr@maths.ox.ac.uk}}
\vskip0.5cm

\normalsize
\medskip

\textit{Mathematical Institute, University of Oxford, \\ Andrew Wiles Building, Radcliffe Observatory Quarter, \\
Woodstock Road, Oxford, OX2 6GG, UK}

\end{center}

\vskip0.6cm

\begin{abstract}

{\begin{center}
    \textbf{Abstract}
\end{center}}

\noindent Carrollian holography is a framework for flat space holography, suggesting that gravity in asymptotically flat spacetime in $D$ dimensions is dual to a conformal Carrollian field theory in $D - 1$ dimensions living at null infinity. In this work, we elaborate on the definition of Carrollian amplitudes for massless scalar fields in general dimensions and provide explicit expressions for the two-, three-, and four-point functions. We show that these amplitudes naturally arise from Lorentzian holographic correlators in AdS/CFT through a correspondence between the flat space limit in the bulk and the Carrollian limit at the boundary. Finally, we use the relation between Carrollian and celestial holography to derive explicit expressions for celestial amplitudes in $D$ dimensions, which are reinterpreted as correlators of the celestial CFT in $D - 2$ dimensions. 
\end{abstract}

\end{titlepage}
\setcounter{page}{2}

\setcounter{tocdepth}{2}
\tableofcontents

\section{Introduction}
\label{sec:intro}
Asymptotically flat spacetimes constitute useful models to describe real-world observable phenomena. For this reason, the flat space holography program so far has mostly focused on bulk spacetimes in four dimensions. For instance, the proposal of celestial holography suggests that gravity in 4d flat space is dual to a 2d CFT living on the celestial sphere (see \cite{Strominger:2017zoo, Pasterski:2021raf} and references therein). This framework was later related to Carrollian holography \cite{Donnay:2022aba, Bagchi:2022emh, Donnay:2022wvx}, which instead suggests that the dual theory is a 3d Carrollian CFT living at null infinity ($\mathscr{I}$). Some Carrollian field theories can be constructed from standard relativistic field theories by implementing a Carrollian limit, i.e. taking the speed of light to zero, $c\to 0$ \cite{Levy1965} (see e.g. \cite{Bagchi:2019xfx, Henneaux:2021yzg,Hansen:2021fxi,Chen:2023pqf,Bergshoeff:2023vfd,Cotler:2025dau} for explicit examples). In this setup, massless scattering amplitudes in flat space can be recast as Carrollian CFT correlators at $\mathscr{I}$, called ``Carrollian amplitudes'' \cite{Donnay:2022wvx,Mason:2023mti} (see also \cite{Banerjee:2019prz,Salzer:2023jqv,Saha:2023hsl,Bagchi:2023fbj,Nguyen:2023miw,Bagchi:2023cen,Ruzziconi:2024zkr,Liu:2024nfc,Have:2024dff,Stieberger:2024shv,Adamo:2024mqn,Alday:2024yyj,Banerjee:2024hvb,Kraus:2024gso,Jorstad:2024yzm,Ruzziconi:2024kzo,Kulp:2024scx,Kraus:2025wgi,Nguyen:2025sqk} for recent developments). Evidence has shown that Carrollian holography naturally arises from AdS/CFT, through the correspondence between the flat limit in the bulk ($\ell \to \infty$) and the Carrollian limit at the boundary ($c \to 0$) \cite{Barnich:2012aw,Barnich:2012xq, Bagchi:2012xr, Ciambelli:2018wre,Compere:2019bua,Compere:2020lrt,Campoleoni:2023fug}. This has recently been exploited in \cite{Alday:2024yyj,Lipstein:2025jfj} to deduce features of a top-down 4d Carrollian hologram and compute flat space scattering amplitudes from an intrinsic boundary computation.

However, from the example of AdS/CFT, we know that explicit realizations of dualities may be more tractable in other bulk dimensions — a concrete example is the celebrated duality between type IIB string theory on AdS$_5 \times S^5$ and $\mathcal{N}=4$ Super-Yang–Mills theory \cite{Maldacena:1997re}. Several works have extended the aforementioned proposal of celestial holography to higher dimensions; see e.g. \cite{Kapec:2015vwa,Kapec:2017gsg,He:2019jjk,He:2019pll,Pano:2023slc,deGioia:2024yne}. One of the main features is that the dual theory is no longer a 2d CFT, and the conformal symmetries form a finite-dimensional group.\footnote{An infinite-dimensional enhancement of the conformal algebra with superrotations would require smooth superrotations \cite{Campiglia:2014yka, Campiglia:2015yka, Compere:2018ylh,Campoleoni:2020ejn,Capone:2023roc}, which erase the CFT structure.} Moreover, in the Carrollian holography proposal, a discussion of Carrollian amplitudes in higher dimensions was provided in \cite{Liu:2024llk} and, for the three-dimensional case, in \cite{Surubaru:2025fmg}.

In this paper, we formulate the Carrollian holography dictionary in general dimension \( D \)$\geq 4$.\footnote{In the rest of the paper, we will always assume $D \geq 4$.}  More precisely, we elaborate on the definition of Carrollian amplitudes that encode bulk scattering amplitudes in \( D \) dimensions in terms of Carrollian CFT correlators at null infinity, and we discuss their relation to the flat space extrapolate dictionary. We provide the explicit form of the two-, three-, and four-point Carrollian amplitudes for massless scalar fields. We then connect this framework with AdS/CFT, showing that Carrollian correlators encoding amplitudes in \( D \) dimensions arise from the flat space/Carrollian limit of holographic CFT\(_{D-1}\) boundary correlators. This extends the correspondence between the flat space limit in the bulk and the Carrollian limit at the boundary of AdS/CFT to \( D \) dimensions. Finally, we revisit the relation between Carrollian and celestial amplitudes in general dimensions and provide explicit expressions for the latter, which we compare with \cite{deGioia:2024yne}. Figure \ref{fig:diag} summarizes the connections we investigate and provides references to the sections containing the corresponding results.

The rest of the paper is organized as follows. In Section~\ref{sec:prelims}, we discuss the basic ingredients for establishing Carrollian holography in \( D \) dimensions, including flat Bondi coordinates, the definition of boundary operators, and the flat-space bulk-to-bulk and bulk-to-boundary propagators. In Section~\ref{sec:carramps}, we define Carrollian amplitudes in \( D \) dimensions and compute the explicit form of the two-, three-, and four-point functions. In Section~\ref{sec:adscorrs}, we discuss the flat space limit of bulk AdS/CFT correlators in position space and show that they naturally give rise to Carrollian amplitudes. In Section~\ref{sec:carrlimitcftcorrs}, we compute the Carrollian limit of the holographic CFT correlators directly at the boundary and recover the Carrollian correlators at null infinity, thereby illustrating the correspondence between the flat space limit and the Carrollian limit of AdS/CFT at the level of correlators. In Section~\ref{sec:From Carrollian to celestial amplitudes}, we discuss the relation between Carrollian and celestial primaries and amplitudes, and derive explicit expressions for the two-, three-, and four-point celestial amplitudes in general dimensions. Finally, in Section~\ref{sec:Outlook}, we conclude with some implications of our work for the flat space holography program.

This paper also contains an Appendix~\ref{sec:ctildecalc}, where we discuss an additional contribution to the three-point function arising in the Carrollian limit of a three-point CFT correlator.

\begin{center}
\begin{figure}[h]
\centering
\begin{tikzpicture}[
  every node/.style={font=\small},
  box/.style={draw, rounded corners, minimum width=3.8cm, minimum height=1cm, align=center},
  arrow/.style={-{Latex}, thick},
  barrow/.style={<->, thick}  
]

\node[box] (witten) at (0,0) {Witten diagrams \\ in $\mathrm{AdS}_D$};
\node[box] (carrollAmp) at (7,0) {Feynman diagrams for \\ Carrollian amplitudes in $D$ dimensions};

\node[box] (holoCFT) at (0,-2.5) {Holographic $\mathrm{CFT}_{D-1}$ \\ correlators};
\node[box] (carrollCFT) at (7,-2.5) {Carrollian $\mathrm{CFT}_{D-1}$ \\ correlators};

\node[box] (celestialCFT) at (7,-5) {Celestial CFT$_{D-2}$ \\ correlators};

\draw[arrow] (witten) -- node[above] {$\ell \to \infty$ (\ref{sec:adscorrs})} (carrollAmp);
\draw[arrow] (holoCFT) -- node[below] {$c \to 0$ (\ref{sec:carrlimitcftcorrs})} (carrollCFT);
\draw[barrow] (holoCFT.south) -- node[pos=0.5, yshift=-4mm] {\cite{deGioia:2024yne}} (celestialCFT.west);
\draw[barrow] (witten) -- node[left] {AdS/CFT} (holoCFT);
\draw[barrow] (carrollAmp) -- node[right] {Carrollian holography (\ref{sec:prelims}, \ref{sec:carramps})} (carrollCFT);
\draw[barrow] (carrollCFT) -- node[right] {Carroll/Celestial correspondence (\ref{sec:From Carrollian to celestial amplitudes})} (celestialCFT);

\end{tikzpicture}
\caption{Summary of the connections investigated in the paper.}
\label{fig:diag}
\end{figure}
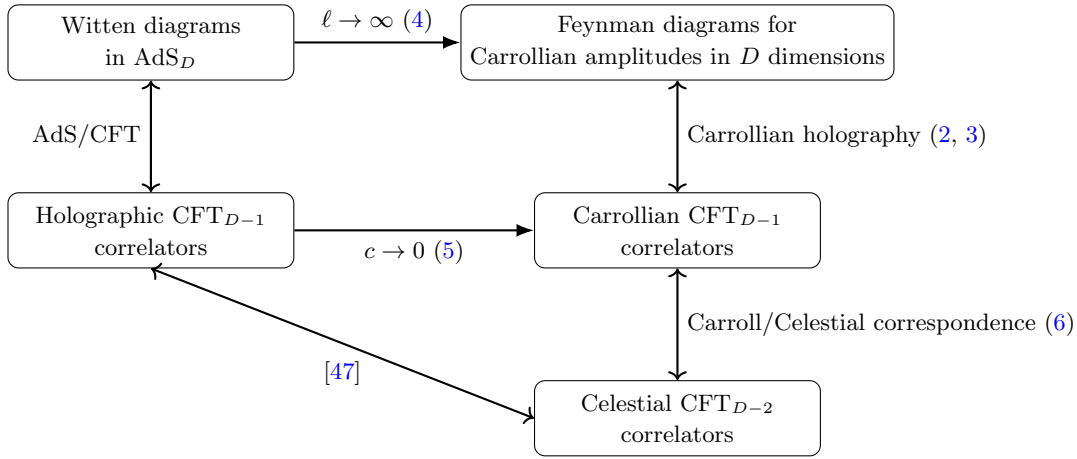
\end{center}

\section{Preliminaries}
\label{sec:prelims}
In this section, we introduce the building blocks of the Carrollian holography dictionary. These include flat Bondi coordinates in $D$ dimensions, the definition and transformation laws of conformal Carrollian primaries, the relation between bulk fields and boundary operators, and the construction of bulk-to-bulk and bulk-to-boundary propagators. This extends the discussion of Section 2 of \cite{Donnay:2022wvx} as well as Section 2 of \cite{Alday:2024yyj} to arbitrary spacetime dimensions $D$.

\subsection{Bondi coordinates}
\label{sec:flatbondi}

In this section, we will work in Minkowski spacetime $\mathbb{R}^{D-1,1}$, on which we will use two coordinate systems, rectangular coordinates $X^{\mu}$ and flat Bondi coordinates $\left(u, r, {\bf x}\right)$, which are related by
\begin{align}
\label{eq:bondicoords}
    X^{\mu} = \frac{u\, \Box_{{\bf x}} q^{\mu} }{D-2} + r\, q^{\mu} = \left(u+\frac{r}{2}\left(1+\left|{\bf x}\right|^2\right), r \,{\bf x}, -u-\frac{r}{2}\left(-1+\left|{\bf x}\right|^2\right)\right).
\end{align}
Here $u, r \in \mathbb{R}$, $ {\bf x} \in \mathbb{R}^{D-2}$ are coordinates on the Celestial Sphere $S^{D-2}$, $\Box_{{\bf x}}$ is the Laplacian in these coordinates and the vector
\begin{align}
    \label{eq:qdef}
    q^{\mu} \left({\bf x}\right)= \frac{1}{2} \left(1+ \left|{\bf x}\right|^2, 2 {\bf x}, 1- \left|{\bf x}\right|^2\right)
\end{align} 
picks out one null direction in $\mathbb{R}^{D}$ for every point on the Celestial sphere $S^{D-2}$. The metric in flat Bondi coordinates is 
\begin{align}
    \label{eq:bondimetricFlat}
    ds_{\mathbb{R}^{D-1,1}}^2 = -2du \, dr+  r^2 \left|d{\bf x}\right|^2.
\end{align}
The null boundaries $\scri^{\pm}$ are reached when the radial coordinate $r \to \pm\infty$. Thus, both the infinite past and future are covered by one set of coordinates. The metric at this conformal boundary is 
\begin{align}
     ds_{\scri^{\pm}}^2 = 0\, du^2+  \left|d{\bf x}\right|^2.
\end{align}
Points on this conformal boundary are denoted by $x = \left(u, {\bf x}\right).$ In these flat Bondi coordinates, we can use the same coordinates at $\mathscr{I}^+$ and $\mathscr{I}^-$: start with a point $\left(u, {\bf x}\right)$ at $\mathscr{I}^-$ situated at $r \to -\infty$, and follow the null geodesic generated by $\partial_r$. You will reach a point at $\mathscr{I}^+$ when $r\to +\infty$ with the same coordinate $\left(u, {\bf x}\right)$. However, despite this geometric identification, it will be crucial to remember if operators are inserted at $\mathscr{I}^+$ or $\mathscr{I}^-$ by using an index $\epsilon = \pm 1$, see e.g. Equation \eqref{eq:boundaryopdef} below. 

\subsection{Conformal Carrollian primaries}
\label{sec:carrprimaries}

We will be interested in the effects of Poincar{\'e} transformations $X^{\mu} \to X^{\prime \mu} = {\Lambda^{\mu}}_{\nu} X^{\nu} + t^{\mu}$ at $\scri^{\pm}$. The boundary coordinates transform as
\begin{align}
    \label{eq:poincaretransform}
    u \to u' = \left|\frac{\partial {\bf x}'}{\partial {\bf x}}\right|^{\frac{1}{D-2}} \left(u - q^\mu({\bf x}) {\Lambda_\mu}^\nu t_\nu \right), \qquad {\bf x} \to {\bf x}',
\end{align}
where ${\bf x} \to {\bf x}'$ is the conformal transformation at the boundary induced by the Lorentz transformation in the bulk.
We will also require the following parametrization of a null momentum of particle $i$: 
\begin{align}
\label{eq:mompar}
    p_i^{\mu} = \e_i \om_i q^{\mu}\left({\bf y}\right), \qquad i = 1, \dots n
\end{align}
where $\e = \pm 1$ for incoming/outgoing particles. Under a Lorentz transformation of the momentum $p^{\mu} \to p^{'\mu} = {\Lambda^{\mu}}_{\nu} p^{\nu}$
\begin{equation}
\label{eq:omegatransformation}
   {\bf y} \to {\bf y}', \qquad \om \to \om' = \left|\frac{\partial {\bf y}'}{\partial {\bf y}} \right|^{-\frac{1}{D-2}} \om.
\end{equation}

The boundary coordinate transformations \eqref{eq:poincaretransform} induced by bulk Poincar\'e transformations can be re-interpreted as global conformal Carrollian transformations at $\mathscr{I}^\pm$ \cite{Duval:2014uva}. Conformal Carrollian primary fields have been studied in general dimensions; see, e.g. \cite{Bagchi:2016bcd,Nguyen:2023vfz}. Here, we restrict to scalar primaries and focus on those that transform trivially under Carrollian boosts, as these are the ones relevant for encoding massless fields \cite{Donnay:2022wvx}. Under \eqref{eq:poincaretransform}, a scalar conformal Carrollian primary field $\Phi_\Delta (u,\bf x)$ with conformal dimension $\Delta$ transforms as
\begin{equation} \label{carrollian primary}
    \Phi_\Delta (u,{\bf x}) \to \Phi'_\Delta (u',{\bf x}') =  \left|\frac{\partial {\bf x}'}{\partial {\bf x}} \right|^{-\frac{\Delta}{D-2}} \Phi_\Delta (u,\bf x)
\end{equation} in $D-1$ dimensions. In the following, we will show that correlators of these primaries at $\mathscr{I}^\pm$ encode the bulk massless scattering amplitudes in $D$ dimensions. Notice that $u-$descendants of conformal Carrollian primaries are also primaries \cite{Donnay:2022wvx}: $\partial_u^m \Phi$ transforms exactly as \eqref{carrollian primary}, but with shifted conformal dimension $\Delta \to \Delta + m$.

\subsection{Boundary operators}
\label{sec:carrboundops}

We can now use the conventions laid out above to define scalar boundary operators and their correlators at $\scri^{\pm}$, extending \cite{Donnay:2022wvx} to general dimensions.\footnote{Spinning boundary correlators can be defined in a similar way.} We start with the Fourier mode expansion of an on-shell massless scalar field in $D$ bulk dimensions: 
\begin{equation}
    \phi(X) = \int \frac{d^{D-1} {\bf p}}{(2\pi)^{D-1} 2p^0} \left[a({\bf p}) e^{ip^\mu X_\mu} + a({\bf p})^\dag e^{-ip^\mu X_\mu} \right] .
\end{equation}
The creation and annihilation operators satisfy the commutation relations
\begin{align}
    \label{eq:commutators}
    &\left[a\left({\bf p}_1\right), a\left({\bf p}_2\right) \right] = \left[a^{\dagger}\left({\bf p}_1\right), a^{\dagger}\left({\bf p}_2\right) \right] = 0, \\
    \nonumber &\left[a\left({\bf p}_1\right), a^{\dagger}\left({\bf p}_2\right) \right] = \left(2\pi\right)^{D-1} \left(2p_1^0\right) \, \delta^{D-1}\left({\bf p}_1 - {\bf p}_2\right) .
\end{align}
Using the parametrization \eqref{eq:mompar} and introducing an $i \varepsilon$ prescription necessary for the integrals to converge, we get
\begin{align}
    \phi(X)= \frac{1}{2(2\pi)^{D-1}} \int_0^\infty d\om \, \om^{D-3} \,d^{D-2}{\bf y}  \left[a(\om, {\bf y}) e^{i\e\om \left(q^\mu X_\mu+i \e\varepsilon\right) } + a(\om, {\bf y})^\dag e^{-i\om \e \left(q^\mu X_\mu-i \e\varepsilon\right)} \right] .
\end{align}
 We are interested in the value of this bulk field at $\scri^{\pm}$. We can approach the boundary in Bondi coordinates \eqref{eq:bondicoords}, in which we have
\begin{align}
    q^\mu \left({\bf y}\right) X_\mu = -u - \frac{r}{2}\left({\bf x}-{\bf y}\right) \cdot \left({\bf x}-{\bf y}\right) \equiv -u- \frac{r}{2}\rho^2,
\end{align}
We have defined ${\bf x}- {\bf y} \equiv \rho {\bf n}$, with ${\bf n}$ being a unit vector on $S^{D-3}$. After this change of variables, we get 
\begin{equation}
    \phi(X) = \frac{1}{2(2\pi)^{D-1}} \int_0^\infty d\om \om^{D-3} \int_0^\infty d\rho \rho^{D-3} \, d^{D-3} {\bf n}\,  e^{-i\e \om r \rho^2} \left[a(\om, {\bf x}-\rho {\bf n}) e^{-i\e \om u-\varepsilon \om} + \text{h.c.}\right].
\end{equation}
As $r \to \e \infty$, we have the saddle point formula
\begin{align}
    \rho^{D-3}\, e^{-i\e \om r \rho^2} \xrightarrow[]{r \to  \e\infty} \frac{2^{\frac{D-4}{2} }}{ \left(i \om\right)^{\frac{D-2}{2}}}\frac{\Gamma \left(\frac{D-2}{2} \right)}{\left|r\right|^{\frac{D-2}{2}}} \delta\left(\rho\right).
\end{align}
Using this and performing the integral over ${\bf n}$ gives:
\begin{equation}
    \phi(X) \xrightarrow[]{r \to \e\infty} \frac{i}{2(2\pi i)^{\frac{D}{2}}} \frac{1}{\left|r\right|^{\frac{D-2}{2}}} \int_0^\infty d \om \,\om^{\frac{D-4}{2}} \left[a(\om, {\bf x}) e^{-i\e \om u-\varepsilon \om} + \text{h.c.}\right] .
\end{equation} The boundary values at $\mathscr{I}^\pm$ of the bulk field is then given by 
\begin{align}
      \lim_{r\to \e \infty} \Big[-2\left(2\pi\right)^{\frac{D-2}{2}}\e^{\frac{D}{2}} \left|r\right|^{\frac{D-2}{2}} \phi(X) \Big] =  \int_0^\infty \frac{d \om}{2\pi} \,\left(-i \e\om\right)^{\frac{D-4}{2}} \left[a(\om, {\bf x}) e^{-i\om u \e-\varepsilon \om} + \text{h.c.}\right]
\end{align} 
where the overall normalization is chosen for convenience. When acting on the in/out vacua, only one of the two terms in the right-hand side will survive. With this in mind, we define the boundary operators
\begin{equation}
\begin{split}
    &\Phi^{+1}(u,{\bf x}) \equiv   \int_0^\infty \frac{d \om}{2\pi} \,\left(-i \e\om\right)^{\frac{D-4}{2}} a\, (\om, {\bf x}) e^{-i\om u -\varepsilon \om},  \\
     &\Phi^{-1}(u,{\bf x}) \equiv   \int_0^\infty \frac{d \om}{2\pi} \,\left(-i \e\om\right)^{\frac{D-4}{2}} a^{\dagger}\, (\om, {\bf x}) e^{i\om u -\varepsilon \om} . 
    \end{split}
\end{equation} or, more compactly, 
\begin{align}
 \label{eq:boundaryopdef}
     &\Phi^{\e}(u,{\bf x})  =  \int_0^\infty \frac{d \om}{2\pi} \,\left(-i \e\om\right)^{\frac{D-4}{2}} a^{\e}\, (\om, {\bf x}) e^{-i\e \om u -\varepsilon \om},
\end{align}
where $\epsilon = \pm 1$ is there to remember if the operator is inserted at $\mathscr{I}^{\pm}$, and $a^{+1} = a$ and $a^{-1} = a^{\dagger}$. A crucial difference compared to the case $D=4$ \cite{Donnay:2022wvx} is that the boundary operator is no longer a Fourier transform of the creation and annihilation operators --- the integrand now includes an additional factor of $\omega^{\frac{D-4}{2}}$. These boundary operators transform as conformal Carrollian primaries under Poincar\'e transformations, which is the key ingredient for the flat space extrapolate dictionary. To see this consider the general Poincar{\'e} transformation described in \eqref{eq:poincaretransform}. Under this, we have 
\begin{equation}
\label{eq:boundaryoptransform1}
  \begin{split}
    \Phi^{\e}\left(u, {\bf x}\right) \to \Phi^{'\e}(u',{\bf x}') &= \int_0^\infty \frac{d\om}{2\pi} \,\left(-i \e\om\right)^{\frac{D-4}{2}}\left[a^\epsilon (\om, {\bf x}') e^{-i\e\om u'-\varepsilon \om} \right],\\
    &= \left|\frac{\partial {\bf x}'}{\partial {\bf x}} \right|^{-\frac{1}{2}} \Phi(u, {\bf x}).    
  \end{split}
\end{equation}
In arriving at the last equality, we changed variables to $\om= \left|\frac{\partial {\bf x}'}{\partial {\bf x}} \right|^{-\frac{1}{D-2}} \om'$, and used the transformation laws for the creation and annihilation operators
\begin{equation}
\label{eq:boundaryoptransform2}
    a'(\om', {\bf x}') = e^{-i\om q^\mu({\bf x}) {\Lambda_\mu}^\nu t_\nu}  a(\om, {\bf x}) \ \ \ \ , \ \ \ \ a^\dag{^\prime}(\om', {\bf x}') = e^{i\om q^\mu({\bf x}) {\Lambda_\mu}^\nu t_\nu}  a^\dag(\om, {\bf x}).
\end{equation}
The final equality is precisely the transformation law for a conformal Carrollian primary \eqref{carrollian primary} with conformal dimension $ \Delta = \frac{D-2}{2}$ in $D-1$ dimensions. In $D=4$, we recover the result that the the boundary values of the bulk fields coincide with conformal Carrollian primaries with $\Delta=1$ \cite{Barnich:2021dta}. The boundary operators act on the in and out vacua to prepare in and out states for scattering amplitudes expressed in a position space basis at $\scri^{\pm}$: 
\begin{align}
    &\Phi^{-1}\left(u, {\bf x}\right) \left|0\right>_{\text{in}} = \int \frac{d\om}{2\pi} \, \left(-i\om\right)^{\frac{D-4}{2}} e^{i\om u -\varepsilon \om} a^{\dagger}\left(\om, {\bf x}\right) \left|0\right>_{\text{in}} = \int \frac{d\om}{2\pi} \, \left(-i\om\right)^{\frac{D-4}{2}}  e^{i\om u-\varepsilon \om} \left|\om, {\bf x}\right>_{\text{in}}\\
    & {}_{\text{out}}\left<0\right|\Phi^{+1}\left(u, {\bf x}\right) = \int \frac{d\om}{2\pi} \, \left(i\om\right)^{\frac{D-4}{2}} e^{-i\om u-\varepsilon \om} {}_{\text{out}}\left<0\right|a\left(\om, {\bf x}\right) \equiv \int \frac{d\om}{2\pi} \, \left(i\om\right)^{\frac{D-4}{2}} e^{-i\om u-\varepsilon \om} {}_{\text{out}}\left<\om, {\bf x}\right|\nonumber,
\end{align}
where $\left|\om, {\bf x}\right>_{\text{in}}$ and ${}_{\text{out}}\left<\om, {\bf x}\right|$ are one particle states in a momentum basis.

\subsection{Propagators}
\label{sec:flatprops}

For the purposes of connecting with the flat limit of AdS Witten diagrams, it is useful to give a definition of Carrollian amplitudes that mirrors that of Witten diagrams, involving bulk-to-bulk and bulk-to-boundary propagators. These are sometimes refered to as the ``Feynman rules for Carrollian amplitudes'' \cite{Liu:2024nfc,Alday:2024yyj}. The bulk-to-bulk propagator for a massless scalar field in flat space is the solution to the differential equation 
\begin{align}
    \label{eq:GBBflatdef}
    \Box_{X_1} G_{BB}^{Flat}\left(X_1, X_2\right) = \left(\frac{\Box_{{\bf x}_1}}{r_1^2}-\frac{D-2}{r}\partial_u-2\partial_u \partial_r\right) G_{BB}^{Flat}\left(X_1, X_2\right) = \delta^{D}\left(X_1-X_2\right),
\end{align}
where $\Box_{X_1}$ is the wave operator in the rectangular coordinates $X_1^{\mu}$ in $D$ dimensions and $\Box_{{\bf x}_1}$ is the wave operator in the $D-2$ dimensional coordinates ${\bf x}_1$. Of course, this coincides with the standard Feynman propagator, which can be easily expressed in momentum space as 
\begin{align}
\label{eq:GBBflat}
    \mathcal{G}^{Flat}_{BB}(X_1,X_2) &= -\int \frac{d^Dp}{(2\pi)^D}\frac{e^{-i p.X_{12}}}{(p^2+i\varepsilon)} = \frac{-i}{4 \pi^{D/2}} \frac{\Gamma(\frac{D-2}{2})}{(X_{12}^2 + i\varepsilon)^{\frac{D-2}{2}}} \nonumber \\
    &=\frac{-i}{4 \pi^{D/2}} \frac{\Gamma(\frac{D-2}{2})}{(-2 r_{12} u_{12} +  r_1r_2\left|{\bf x}_{12}\right|^2 + i\varepsilon)^{\frac{D-2}{2}}}.
\end{align}
In the second line, we have expressed the bulk-to-bulk propagator in position space in Bondi coordinates \eqref{eq:bondicoords} in which $X_{12}^2 = -2 r_{12} u_{12} + r_1r_2 \sum_{i=1}^{D-2} x_i^2$.
We can obtain the bulk-to-boundary propagator by sending one of the points to $\scri^{\pm}$:
\begin{align}
\label{eq:GBbflat}
    \mathcal{G}^{Flat}_{Bb,\pm}(x_1,X_2) &= \lim_{r_1\to \pm\infty} \left|r_1\right|^{\frac{D-2}{2}} \mathcal{G}^{Flat}_{BB}(X_1,X_2) \nonumber \\
    &= \frac{-i}{2\left(2\pi\right)^{D/2}} \frac{\Gamma(\frac{D-2}{2})}{(-u_1 - q_1.X_2 \pm i\epsilon)^{\frac{D-2}{2}}} .
\end{align}
It will be useful for us to rewrite the bulk-to-boundary propagator as an integral transform:
\begin{align}
    \label{eq:inttransform}
    \mathcal{G}^{Flat}_{Bb,\pm}(x_1,X_2)  = \frac{-\epsilon}{2(2\pi)^{D/2}} \int_0^\infty d\omega (-i\epsilon \omega)^{\frac{D-4}{2}} e^{-i\epsilon \omega u_1 - i\epsilon\omega q_1.X_2} e^{-\varepsilon \omega}.
\end{align}
It will also be useful to consider the bulk-to-boundary propagators for $u-$descendants,
\begin{equation}\label{eq:GBbudesc}
\begin{split}
\mathcal{G}^{Flat,\D}_{Bb,\pm}(x_1,X_2)  \equiv  \partial_{u_1}^{m}\mathcal{G}^{Flat}_{Bb,\pm}(x_1,X_2)   &= \frac{-i}{2 \left(2\pi\right)^{\frac{D}{2}}} \frac{\Gamma\left(\frac{D-2}{2}+m\right)}{(-u_1 - q_1.X_2 \pm i\epsilon)^{\frac{D-2}{2}+m}} \\
&= \frac{-i}{2 \left(2\pi\right)^{\frac{D}{2}}} \frac{\Gamma\left(\D\right)}{(-u_1 - q_1.X_2 \pm i\epsilon)^{\Delta}} .
\end{split}  
\end{equation} In the last line, we have rewritten the formula in terms of the conformal weight
\begin{equation}
    \D = \frac{D-2}{2}+m.
\label{conformal weight}
\end{equation} We can also give an integral representation for these bulk-to-boundary propagators:
\begin{align}
\mathcal{G}^{Flat,\D}_{Bb,\pm}(x_1,X_2)  &= \frac{-\epsilon}{2(2\pi)^{D/2}}\int_0^\infty d\omega (-i\epsilon \omega)^{m+\frac{D-4}{2}} e^{-i\epsilon \omega u_1 - i\epsilon\omega q_1.X_2} e^{-\varepsilon \omega} \\
      &= \frac{-\epsilon}{2(2\pi)^{D/2}}\int_0^\infty d\omega (-i\epsilon \omega)^{\D-1} e^{-i\epsilon \omega u_1 - i\epsilon\omega q_1.X_2} e^{-\varepsilon \omega}.\nonumber
\end{align}

\section{Carrollian amplitudes in $D$ dimensions}
\label{sec:carramps}

In this section, we discuss the definition of Carrollian amplitudes in general dimensions \cite{Liu:2024llk}, which extends the definition given in \cite{Donnay:2022wvx,Mason:2023mti,Alday:2024yyj} in the $D=4$ case. The idea is to recast the usual bulk amplitudes in terms of correlators at $\mathscr{I}^{\pm}$ which can then be re-interpreted as Carrollian correlators in a putative dual theory. We give explicit expressions of the latter for the two-, three- and four-point functions for a massless scalar field.

\subsection{Definitions}

We will now give two equivalent definitions of Carrollian amplitudes in $D$ dimension. First, based on \cite{Donnay:2022wvx}, we can define them as correlators of the boundary operators defined in \eqref{eq:boundaryopdef}:
\begin{align}
    \mathcal{C}_n \left(u_i, {\bf x}_i \right) &= {}_{\text{out}}\an{0| \Phi^{\e_1}\left(u_1, {\bf x}_1 \right) \dots \Phi^{\e_n}\left(u_n, {\bf x}_n \right)|0}_{\text{in}} \nonumber \\ 
    &=\,  \int_{0}^{\infty} \prod_{k=1}^n \frac{d\om_k}{2\pi}  \,\left(-i \e_k \om_k\right)^{\frac{D-4}{2}}  e^{-i \e_k u_k \om_k-\varepsilon \om_k} \mathcal{A}_n \left(\om_i, {\bf x}_i \right). 
    \label{eq:carrampdefbound}
\end{align}
We have left it implicit in the above equation that the boundary operators with $\e=\pm 1$ act on the ``out'' and ``in''  vacua respectively. $\mathcal{A}_n \left(\om_i, {\bf x}_i \right)$ is the corresponding momentum space scattering 
\begin{align}
      \mathcal{A}_n \left(\om_i, {\bf x}_i \right) = {}_{\text{out}}\an{\left\lbrace \om_1, {\bf x}_1\right\rbrace, \dots \left\lbrace \om_k, {\bf x}_k \right\rbrace |\left\lbrace \om_{k+1}, {\bf x}_{k+1}\right\rbrace, \dots \left\lbrace \om_n, {\bf x}_n\right\rbrace}_{\text{in}},
\end{align}
where we have assumed that operators $1, \dots ,k$ are outgoing and $k+1, \dots , n$ are incoming. Notice that \eqref{eq:carrampdefbound} is no longer a Fourier transform in $D \neq 4$ because of the prefactor in the integrand, see also \eqref{eq:boundaryopdef} and the text below. Under a conformal Carrollian transformation at $\mathscr{I}^\pm$ \eqref{eq:poincaretransform}, we have 
\begin{equation} \label{Carroll1}
    \mathcal{C}'_n \left(u'_i, {\bf x}'_i \right) = \prod_{i=1}^n\left|\frac{\partial {\bf x}'_i}{\partial {\bf x}_i} \right|^{-\frac{1}{2}} \times \mathcal{C}_n \left(u_i, {\bf x}_i \right)
\end{equation}
since each boundary operator is a conformal Carrollian primary with weight $\D_k = \frac{D-2}{2}$. It is therefore natural to identify Carrollian amplitudes \eqref{eq:carrampdefbound} with correlators in the dual theory which satisfy Carrollian CFT Ward identities. 

Since $u-$descendants of conformal Carrollian primaries are also primaries (see below \eqref{carrollian primary}), it is natural to consider their correlators:
\begin{align} \label{descendant Carrollian amplitude}
      \mathcal{C}_n^{\D_1, \dots , \D_n} \left(u_i, {\bf x}_i \right)  \equiv \prod_{k=1}^n\partial_{u_k}^{m_k}  \mathcal{C}_n\left(u_i, {\bf x}_i \right)  &= \int_{0}^{\infty} \prod_{k=1}^n \frac{d\om_k}{2\pi}  \,  e^{-i \e_k u_k \om_k}  \left(-i\e_k \om_k\right)^{\frac{D-4}{2}+m_k}\, \mathcal{A}_n \left(\om_i, {\bf x}_i \right)\\
      &\nonumber =\int_{0}^{\infty} \prod_{k=1}^n \frac{d\om_k}{2\pi}  \,  e^{-i \e_k u_k \om_k}  \left(-i\e_k \om_k\right)^{\D_k-1}\, \mathcal{A}_n \left(\om_i, {\bf x}_i \right),
\end{align}
where in the final line, we have written the formula in terms of the conformal weights $\D_k = m_k + \frac{D-4}{2}$, see \eqref{conformal weight}. As usual, one can extend this definition for any $\Delta \in \mathbb{C}$ via the modified Mellin transform \cite{Banerjee:2018gce,Banerjee:2018fgd}. 

The second definition of Carrollian amplitudes is based on Feynman diagrams in position space and is closely related to the definition of AdS boundary correlators via Witten daigrams. Given a collection of Feynman diagrams contributing to the ampltiude, to every external line, we associate a bulk-to-boundary propagator \eqref{eq:GBbflat} and for every internal line, the bulk-to-bulk propagator \eqref{eq:GBBflat}. Integrating over the bulk points, we are left with a function of the external positions. Using the integral representation \eqref{eq:inttransform}, we can relate this to the definition \eqref{eq:carrampdefbound}. We will demonstrate this with explicit examples in the following sections. 

\subsection{Two-point amplitude}
\label{sec:2ptcarramps}

We will first compute Carrollian amplitudes of massless scalars in a $D$-dimensional bulk spacetime using the definition \eqref{eq:carrampdefbound}. We start with the two-point amplitude in momentum space, which is 
\begin{align}
    \mA_2 = 2 \kappa_2 p_1^0 \delta^{D-1} \left(p_1+p_2\right) =\frac{2\kappa_2}{\om_1^{D-3}} \delta^{D-2}\left({\bf x}_{12}\right) \delta\left(\om_1-\om_2\right)\delta_{\e_1, -\e_2},
\end{align}
where we used the parametrization \eqref{eq:mompar} and introduced the notation ${\bf x}_{12} \equiv {\bf x}_1 - {\bf x}_2$. The corresponding Carrollian amplitude is 
\begin{align}
\label{eq:2ptcarramp}
    \mc_2^{\D_1, \D_2} &=\int_0^{\infty} \prod_{k=1}^2\frac{d\om_k}{2\pi} \left(-i\e_k \om_k\right)^{\D_k-1} e^{-i \e_k u_k \om_k -\varepsilon \om_k} \mA_2  \\
    &\nonumber =\frac{\kappa_2}{2\pi^2} i^{D} (-1)^{\D_1}\frac{\delta^{D-2}\left({\bf x}_{12}\right) \Gamma\left(\Sigma_{\D}-(D-2)\right)}{\left(u_{12}-i \e_1 \varepsilon\right)^{\Sigma_{\D}-(D-2)}} \delta_{\e_1, -\e_2},
\end{align}
with $\Sigma_{\D}=  \D_1+\D_2$. Notice that the integral diverges if $\D_1 = \D_2 = \frac{D-2}{2}$, as in the $D = 4$ case \cite{Donnay:2022wvx}.

\subsection{Thee-point amplitude}
\label{sec:3ptcarramps}

The three-point amplitude for scalars is
\begin{align}
    \mA_3 = \kappa_3\, \delta^{D}\left(\sum_{k=1}^3 p_k\right) = \frac{\pi^{\frac{D-2}{2}}}{\Gamma\left(\frac{D-2}{2}\right)}\frac{\kappa_3 \left|{\bf x}_{12}\right|^{D-4}}{\om_1 \om_2 \om_3^{D-3}} \,\delta\left(\sum_{k=1}^3 \e_k \om_k\right) \delta^{D-2}\left({\bf x}_{12}\right) \delta^{D-2}\left({\bf x}_{23}\right).
\end{align}
It is important to point out that the second equality only holds in Lorentzian signature, where all three particles must be collinear. Such amplitudes have also been considered in \cite{Bagchi:2023fbj, deGioia:2024yne}. The corresponding Carrollian amplitude is: 
\begin{align}
\label{eq:3ptcarrampdef1}
    \mc_3^{^{\D_1, \D_2, \D_3}} &= \kappa_3\int \prod_{k=1}^3 \frac{d\om_k}{2\pi}\left(-i\e_k \om_k\right)^{\D_k-1} e^{-iu_k\om_k \e_k} \delta^{D}\left(\sum_{j=1}^3 p_j\right)
\end{align}
We will set  $-\e_1=\e_2=\e_3=1$ for concreteness. In this case, we have:
\begin{align}
     \mc_3^{^{\D_1, \D_2, \D_3}}  &=\nonumber  \frac{\kappa_3}{8}\frac{\pi^{\frac{D-8}{2}}}{\Gamma\left(\frac{D-2}{2}\right)}\left(-i\right)^{\Sigma_{\D}-3}\, \left|{\bf x}_{12}\right|^{D-4} \delta^{D-2}\left({\bf x}_{12}\right) \delta^{D-2}\left({\bf x}_{23}\right) \\
    &\qquad\times \int d\om_2 d\om_3  \left(\om_2+\om_3\right)^{\D_1-2} \om_2^{\D_2-2} \om_3^{\D_3-D+2} e^{i \left(u_{12}\om_2+u_{13}\om_3\right)}
\end{align}
where $\Sigma_{\D} = \sum_{i=1}^3 \D_i$. After performing the integral, we get
\begin{align}
    \label{eq:3ptcarramp}
    \mc_3^{\D_1, \D_2, \D_3} &=   \frac{\kappa_3}{8} \pi^{\frac{D-8}{2}}\left(-i\right)^{D-3}\,\frac{\Gamma\left(\Sigma_\D - D\right) B(\D_2 - 1, \D_3 - D+3)}{\Gamma\left(\frac{D-2}{2}\right)}  \left|{\bf x}_{12}\right|^{D-4} \delta^{D-2}\left({\bf x}_{12}\right) \delta^{D-2}\left({\bf x}_{23}\right) \nonumber\\
    &\qquad \times u_{12}^{D-\Sigma_{\D}} 
    \hypf\left(\Sigma_{\D}-D,\D_3-(D-3),\D_2 - 1;\frac{u_{32}}{u_{12}}\right).
\end{align}
Note that in doing the integral we have used that the Gauss hypergeometric function $\hypf(a,b,c;z)$ has an integral representation only when Re$(z) < 1$, which translates to the choice of ordering $u_{32} < u_{12}$. 
We can also obtain this from the second, Feynman diagrammatic definition of the three-point Carrollian amplitude:
\begin{align}
\label{eq:3ptfeyndiagdef}
    \mc_3^{\D_1, \D_2, \D_3} & \, =\, \beta_3 \kappa_3 \int d^DX   \prod_{i=1}^3\mathcal{G}^{Flat,m_i}_{Bb,\e_i}({\bf x}_i,X) \\
    &\nonumber = -\frac{\e_1 \e_2 \e_3\beta_3 \kappa_3}{8(2\pi)^{3\frac{D}{2}}} \int d^DX   \int_0^{\infty}\prod_{k=1}^3 d\om_k \left(-i \e_k \om_k\right)^{\D_k-1} e^{-i\sum_{k=1}^3\left( \e_k u_k \om_k+\e_k\om_K q_k \cdot X-\varepsilon \om_k\right)}.
\end{align}
In arriving at this, we have used the integral representation of the bulk-to-boundary propagators \eqref{eq:inttransform}. Performing the integral over $X$ to get $\left(2\pi\right)^D \delta^D \left(\sum_{i=3}^3 p_i\right)$ and comparing with \eqref{eq:3ptcarrampdef1}, we find an equality if $\beta_3 = -\frac{\left(2\pi\right)^{D/2}}{\pi^3}\e_1\e_2\e_3$.

\subsection{Four-point amplitude}
\label{sec:4ptcarramp}

We will focus on the contact $4$-point diagram between scalars (as discussed in \cite{Alday:2024yyj} for the $D=4$ case, this analysis can be easily extended to exchange diagrams). The momentum space amplitudes for this is
\begin{align}
\label{eq:4ptmomspace}
    \mA_{4} = \kappa_{4} \,\delta^{D}\left(P^{\mu}\right),
\end{align}
where $P^{\mu} = \sum_{i=1}^4 \om_i \e_i q_i^{\mu}$ is the total momentum. We will first compute the Carrollian amplitude by using the above formula and \eqref{descendant Carrollian amplitude}:
\begin{align}
    \label{eq:4ptcarrramp}
    C_{4}^{\D_1, \dots \D_4} = \int \prod_{k=1}^4 \frac{d\om_k}{2\pi} \left(-i\e_k \om_k\right)^{\D_k-1} e^{-iu_k \e_k \om_k} \mA_4.
\end{align}
In order to perform this integral, we must first solve the equations for momentum conservation. To this end,\footnote{Recall that $D \geq 4$. The decomposition \eqref{eq:4ptdeltafunction1} will only be valid under this assumption.} let $\left(n_5, \dots n_D\right)$ be $D-4$ linearly independent vectors in $\mathbb{R}^{D-1,1}$ that complete $\left(q_1, q_2, q_3, q_4\right)$ into a basis. We can also assume without any loss of generality that 
\begin{align}
\label{eq:constraints}
n_i \cdot q_{j} = 0, \qquad \forall i=5, \dots , D  \text{ and }  j=1,2,3.    
\end{align}
Here and in everything that follows below, the dot product represents a contraction carried out with using the Minkowski metric. We decompose the $\delta$ function as follows:
\begin{align}
\label{eq:4ptdeltafunction1}
    \delta^{D}\left(P^{\mu}\right) &= \det\left(q_1, \dots , q_4, n_5, \dots ,n_D\right) \prod_{i=1}^4 \delta\left(P\cdot q_i\right)\prod_{j=5}^D \delta\left(P \cdot n_j\right), \\
    &\nonumber =\frac{\det\left(q_1, \dots , q_4, n_5, \dots , n_D\right)}{\om_4^{D-3}\left|{\bf x}_{13}\right|^4\left|{\bf x}_{24}\right|^4} \delta\left((z-\zb)^2\right)\prod_{i=1}^3 \delta\left(\om_i - \om_i^{\star}\right)\prod_{j=5}^D \delta\left(q_4 \cdot n_j\right).
\end{align}
We have solved 4 out of the $D$ equations $P^{\mu}=0$ by considering its components along $q_1, \dots , q_4$. The first three give:
\begin{align}
\label{eq:omegastars}
    \om_1^{\star} = -\frac{\left|{\bf x}_{24}\right|^2}{\left|{\bf x}_{12}\right|^2}  z  \e_1 \e_4 \om_4, \quad   \om_2^{\star} = \frac{\left|{\bf x}_{34}\right|^2}{\left|{\bf x}_{23}\right|^2}\frac{1-z}{z}\e_2 \e_4 \om_4, \quad   \om_3^{\star} = -\frac{\left|{\bf x}_{14}\right|^2}{\left|{\bf x}_{13}\right|^2}\frac{1}{1-z}\e_3 \e_4 \om_4,
\end{align}
while the last simply gives $\delta\left((z-\zb)^2\right)$. $z, \zb$ are defined intrinsically  via
\begin{align}
\label{eq:celspherecrossratios}
    z \zb = \frac{\left|{\bf x}_{12}\right|^2\left|{\bf x}_{34}\right|^2}{\left|{\bf x}_{13}\right|^2\left|{\bf x}_{24}\right|^2}, \qquad  (1-z)(1-\zb) = \frac{\left|{\bf x}_{14}\right|^2\left|{\bf x}_{23}\right|^2}{\left|{\bf x}_{13}\right|^2\left|{\bf x}_{24}\right|^2}.
\end{align}
The remaining components $P \cdot n_j$ simply reduce to $q_4 \cdot n_j$ due to \eqref{eq:constraints}. We can further simplify the determinant on the support of the $\delta$ function constraints. 
\begin{align}
   \left| \det\left(q_1, \dots , q_4, n_5, \dots , n_D\right)\right|\prod_{j=5}^D \delta \left(q_4 \cdot n_j\right) &=
\left|
\begin{array}{c|c}
\begin{array}{cccc}
q_1^1 & q_2^1 & q_3^1 & q_4^1 \\
\vdots & \vdots & \vdots & \vdots \\
q_1^4 & q_2^4 & q_3^4 & q_4^4 \\
\end{array}
&
\begin{array}{c}
0
\end{array} \\
\hline
\begin{array}{c}
0
\end{array}
&
\begin{array}{cccc}
n^5_5 & \hdots& & n_5^D \\
\vdots & \vdots & \vdots & \vdots \\
n_D^5 & \hdots & & n_D^D
\end{array}
\end{array}
\right| \prod_{j=5}^D \delta \left(q_4 \cdot n_j\right) \\
& = \frac{1}{4}\left(z-\zb\right){\bf x}_{13}^2 {\bf x}_{24}^2 \det\left(\bar{n}_5, \dots , \bar{n}_D\right)\prod_{j=5}^D \delta \left(q_4 \cdot n_j\right) 
\end{align}
The $D-4$ $\delta$-function constraints $\delta\left(q_4 \cdot n_j\right)$ reduce the full determinant to the determinant of a block-diagonal matrix with $\bar{n}_j$ being the ($D-4$)-dimensional vectors in the directions orthogonal to $q_1, \dots , q_4$. In addition, note that although this vanishes when $z=\zb$, when combined with $\delta((z-\bz)^2)$, this gives a non-trivial result in general since $(z-\zb)\delta\left((z-\zb)^2\right) = \frac{1}{2}\delta\left(z-\zb\right)$ is non-vanishing on the support of $z = \bar z$. All in all, the momentum conserving $\delta$ function reduces to: 
\begin{align}
\label{eq:4ptdeltafunction}
    \delta^{D}\left(P^{\mu}\right) &=\frac{\det\left(\bar{n}_5, \dots , \bar{n}_D\right)}{2\om_4^{D-3}\left|{\bf x}_{13}\right|^2\left|{\bf x}_{24}\right|^2} \delta\left(z-\zb\right)\prod_{i=1}^3 \delta\left(\om_i - \om_i^{\star}\right)\prod_{j=5}^D \delta\left(q_4 \cdot n_j\right).
\end{align}
Without loss of generality, we can choose the ($D-4$)-dimensional vectors $\bar{n}_j$ with $j = 5, \ldots , D$ to be orthonormal, which then implies $\det\left(\bar{n}_5, \dots , \bar{n}_D\right) = 1$. With this, it is straightforward to evaluate the integrals to find a closed-form expression for the four-point Carrollian amplitude in $D$ dimensions. The final result is given by the following compact expression
\begin{align} \label{Carrollian 4pt function}
    \mc_{4}^{\D_1, \dots \D_4} &= \frac{\kappa_4}{8\left(2\pi\right)^4} \frac{\mathcal{S}\,\mathcal{Z}\left({\bf x}_{ij}\right)}{\mathcal{U}^{\Sigma_{\D}-D}}\delta\left(z-\zb\right)\prod_{j=5}^D \delta\left(q_4 \cdot n_j\right) \\ 
  &\qquad\times \left(-1\right)^{\D_1+\D_3-D} z^{\D_1-\D_2} \left(1-z\right)^{\D_2-\D_3} \Gamma\left(\Sigma_{\D}-D\right)\nonumber ,
\end{align}
with the various quantities entering the equation above defined below:
\begin{align}
    &\mathcal{S} =  \Theta\left(-z \e_1 \e_4\right) \Theta\left(\left(1-z\right)z\e_2 \e_4 \right) \Theta\left( (z-1) \e_3 \e_4 \right),\label{eq:Sdef}\\
    &\mathcal{Z}\left({\bf x}_{ij}\right)= \frac{\left|{\bf x}^2_{14}\right|^{\D_3-1}\left|{\bf x}^2_{24}\right|^{\D_1-2}\left|{\bf x}^2_{34}\right|^{\D_2-1}}{\left|{\bf x}^2_{12}\right|^{\D_1-1}\left|{\bf x}^2_{13}\right|^{\D_3}\left|{\bf x}^2_{23}\right|^{\D_2-1}} \label{eq:Zdef},\\
    &\mathcal{U} = u_4 - u_1 z \frac{\left|{\bf x}_{24}\right|^2}{\left|{\bf x}_{12}\right|^2}+u_2 \frac{1-z}{z}\frac{\left|{\bf x}_{34}\right|^2}{\left|{\bf x}_{23}\right|^2} - u_3\frac{1}{1-z}\frac{\left|{\bf x}_{14}\right|^2}{\left|{\bf x}_{13}\right|^2}\label{eq:Udef}.
\end{align}
Note that we can express $\mathcal{U}$ in a manifestly translation invariant way (with $u_{ij} \equiv u_i - u_j$) as
\begin{align}
    \mathcal{U} = - u_{14} z \frac{\left|{\bf x}_{24}\right|^2}{\left|{\bf x}_{12}\right|^2}+u_{24} \frac{1-z}{z}\frac{\left|{\bf x}_{34}\right|^2}{\left|{\bf x}_{23}\right|^2} - u_{34}\frac{1}{1-z}\frac{\left|{\bf x}_{14}\right|^2}{\left|{\bf x}_{13}\right|^2}.
\end{align}
The amplitude can also be defined using bulk-to-boundary propagators as
\begin{align}
\label{eq:4ptfeyndiagdef}
 \mc_{4}^{\D_1, \dots \D_4} & =\, \beta_4 \kappa_4 \int d^DX   \prod_{i=1}^4\mathcal{G}^{Flat,m_i}_{Bb,\e_i}({\bf x}_i,X) \\
    &\nonumber = \frac{\e_1 \e_2 \e_3\e_4 \beta_4 \kappa_4}{16(2\pi)^{2D}} \int d^DX   \int_0^{\infty}\prod_{k=1}^4 d\om_k \left(-i \e_k \om_k\right)^{\D_k-1} e^{-i\sum_{k=1}^4\left( \e_k u_k \om_k+\e_k\om_K q_k \cdot X-\varepsilon \om_k\right)}.
\end{align}
We get an agreement with \eqref{eq:4ptcarrramp} if we choose $\beta_4 = 16 \left(2\pi\right)^{D-4}\e_1\e_2\e_3\e_4.$

\section {AdS${}_D$ boundary correlators and their flat limits}
\label{sec:adscorrs}
In this section, we extend the analysis of \cite{Alday:2024yyj} to general dimensions. For recent and related discussions on the flat space limit of holographic correlators in $D=4$, we also refer to \cite{deGioia:2022fcn,deGioia:2023cbd,deGioia:2024yne,Bagchi:2023fbj, Bagchi:2023cen, Marotta:2024sce}. After reviewing Bondi coordinates and propagators in AdS, we compute the bulk flat space limit of the $2$-, $3$- and $4$-point holographic correlators from the AdS Witten diagram expressions. We will naturally recover the Feynman diagram expressions for Carrollian amplitudes presented in the previous section.

\subsection{Bondi coordinates}
\label{sec:adsbondi}

\paragraph{Bulk metric:}\textit{Lorentzian} AdS${}_D$ is the hyperboloid 
\begin{equation}
    \label{eq:LAdSdef}
    \chi \cdot \chi = -\chi^{+}\chi^{-} -(\chi^0)^2 +\left|\bm{\chi}\right|^2= -\ell^2,
\end{equation} 
where $\chi^I = \left(\chi^+, \chi^-, \chi^0, \bm{\chi}\right)$  are coordinates on the embedding space in $\mathbb{R}^{D-1,2}$ with metric
\begin{align}
    \label{eq:embeddingspacemetric}
ds^2_{\mathbb{R}^{D-1,2}}=  -d\chi^+d\chi^- -(d\chi^0)^2 + \left|d\bm{\chi}\right|^2.
\end{align}
Bondi coordinates $\left(u, r, {\bf x}\right)$ are a set of intrinsic coordinates given by
\begin{align}
    \label{eq:EtoBcoordinates}
    \chi^{I} =  r\left(1, \frac{2u}{r}-\frac{u^2}{\ell^2}+ \left|\bf{x}\right|^2, -\frac{\ell}{r}+\frac{u}{\ell},\bf{x} \right) \ .
\end{align}
It can be checked that these coordinates satisfy the constraint \eqref{eq:LAdSdef}. The AdS metric in Bondi coordinates becomes \cite{Barnich:2012aw, Poole:2018koa, Compere:2019bua,Geiller:2022vto}
\begin{align}
    \label{eq:bondimetricAdS}
    ds^2_{\text{AdS}_D} = -\frac{r^2}{\ell^2} du^2-2 du \, dr +r^2 \left|d{\bf x}\right|^2.
\end{align}
The flat limit is the limit in which $\frac{r}{\ell} \to 0$. In situations where there are multiple distance scales, this is the limit in which every distance scale is smaller than the AdS radius.  Keeping this in mind, in the rest of the paper, we will tacitly denote this limit as $\ell \to \infty$. In this limit, we recover the metric on $D$-dimensional Minkowski spacetime as seen below:
\begin{align}
    \label{eq:flatlimitbondimetric}
     \lim_{\ell \to \infty} ds^2_{\text{AdS}_D} =-2 du \, dr +r^2 \left|d{\bf x}\right|^2 = ds^2_{\mathbb{R}^{D-1,1}}.
\end{align}
We will now discuss various features of this spacetime in Bondi coordinates. These will be useful in later sections in discussing boundary correlators and their flat limits.
\paragraph{Geodesic distance:}The geodesic or chordal distance $\chi_{12}$ between two points $\chi_1, \chi_2$ in AdS is simply the distance between them in embedding space 
\begin{align}
\label{eq:chordaldist}
    \chi_{12} = -r_1 r_2\frac{u_{12}^2}{\ell^2}-2r_{12}u_{12}+r_1 r_2 \left|{\bf x}_{12}\right|^2.
\end{align}
In the flat limit, this reduces to the geodesic distance in Minkowski spacetime in Bondi coordinates as seen from the metric \eqref{eq:bondimetricFlat}. 
\begin{align}
    \lim_{\ell \to \infty}\chi_{12} = -2r_{12}u_{12}+r_1 r_2 \left|{\bf x}_{12}\right|^2 = X_{12}^2.
\end{align}
\paragraph{Conformal boundary:}The conformal boundary of Lorentzian AdS${}_{D}$ is $\mathbb{R}^{D-2,1}$. Points on this boundary, denoted as $\bar{X}^I$ (these are the embedding coordinates for the boundary $\mathbb{R}^{D-2,1}$), are reached by sending $r \to \pm \infty$ and performing a suitable rescaling:
\begin{align}
    \lim_{r \to \pm \infty}\frac{\chi^I}{r} = \bar{X}^I =  \left(1, -\frac{u^2}{\ell^2}+ \left|\bf{x}\right|^2, \frac{u}{\ell},\bf{x} \right) \ .
\end{align}
It is easily seen that $\bar{X}^2 = 0$. Furthermore, because $\bar{X}$ lies on the conformal boundary, it satisfies $\bar{X} \sim \lambda \bar{X}$ for $\lambda\in \mathbb{R}^+$ showing that it is a point in $\mathbb{R}^{D-2,1}$. The metric at the conformal boundary is 
\begin{align}
\label{eq:boudnarymetricads}
    ds^2_{\partial AdS_{D}} =  -\frac{1}{\ell^2} du^2 +\left|d{\bf x}\right|^2.
\end{align}
Note that this is just the Minkowski metric with $\frac{1}{\ell}$ playing the role of the speed of light. The flat limit in the bulk ($\ell \to \infty$) is thus implemented as a Carrollian limit at the boundary and we get
\begin{align}
    \label{eq:flatlimitboundarymetric}
   \lim_{\ell \to \infty} ds^2_{\partial AdS_{D}} =  0\, du^2 +\left|d{\bf x}\right|^2 = ds^2_{\scri^{\pm}},
\end{align}
which is the degenerate metric at the boundary of Minkowski spacetime defining the Carrollian geometry.

\paragraph{Euclidean AdS${}_D$:}We will also use \textit{Euclidean} AdS${}_D$ which is defined as the quadric
\begin{equation}
    \label{eq:EAdSdef}
    \chi \cdot \chi = -\chi^{+}\chi^{-} +(\chi^0)^2 +\left|\bm{\chi}\right|^2= -\ell^2,\ , \qquad \chi^+ + \chi^- >0 \ ,
\end{equation}
in the embedding space $\mathbb{R}^{D,1}$. This is related to the Lorentzian one by analytic continuation $\chi^0 \to i \chi^0$ and has the metric
\begin{align}
    \label{eq:EuclideanAdS}
    ds^2_{\mathbb{R}^{D,1}}=  -d\chi^+d\chi^- +(d\chi^0)^2 + \left|d\bm{\chi}\right|^2.
\end{align}

\subsection{Propagators}
\label{sec:adsprops}

\paragraph{Bulk-to-bulk propagator:} The bulk-to-propagator for a massive scalar with dimension $\D$ in Lorentzian AdS${}_D$ is the solution to the differential equation
\begin{align}
    \label{eq:GBBAdSdef}
   &\left[\nabla^2_{\chi_1}  + m^2\right]G_{BB}^{AdS,\D}\left(\chi_1, \chi_2\right)= \frac{1}{\sqrt{-g}}\delta^D\left(\chi_{12}\right),
\end{align}
where $\nabla^2_{\chi_1} $ is the wave operator on AdS${}_D$ and $m^2\ell^2 = \D\left(D-1-\D\right)$. This can be written in Bondi coordinates as
\begin{align}
    \label{eq:GBBAdSdefbondi}
     \left[\frac{D \,r_1}{\ell^2}\partial_{r_1}+\frac{r^2}{\ell^2}\partial_{r_1}^2+\frac{\Box_{{\bf x}_1}}{r_1^2}-\frac{D-2}{r}\partial_u-2\partial_u \partial_r  + m^2\right]  G_{BB}^{AdS,\D}\left(\chi_1, \chi_2\right)= \frac{1}{\sqrt{-g}}\delta^D\left(\chi_{12}\right)
\end{align} where $\Box_{{\bf x}}$ is the wave operator on $\mathbb{R}^{D-2}$. Notice that as $\ell \to \infty$ and keeping $\Delta$ fixed, the above equation reduces to the massless wave equation in flat space \eqref{eq:GBBflatdef}. It is useful to introduce the following dimensionless parameter related to the chordal distance \eqref{eq:chordaldist}:
\begin{align}
    \rho_{12} = -\frac{4\ell^2}{\chi_{12}} = -\frac{4\ell^2}{-r_1 r_2\frac{u_{12}^2}{\ell^2}-2r_{12}u_{12}+r_1 r_2 \left|{\bf x}_{12}\right|^2}.
\end{align}
Upon changing variables to $\rho_{12}$, we get an ordinary differential equation in one variable. Thus, we need to solve: 
\begin{align}
  \left[\left(4-2D+\rho_{12}\left(D-4\right)\right)\partial_{\rho_{12}} + 2\rho_{12}\left(1-\rho_{12}\right)\partial_{\rho_{12}}^2 +m^2\right] \mathcal{G}_{BB}^{AdS, \D}\left(\rho_{12}\right) = \frac{1}{\sqrt{-g}}\delta^D\left(\chi_{12}\right).
\end{align}
This is a hypergeometric differential equation whose solutions is\footnote{We have discarded the solution which doesn't behave as $\rho_{12}^{\D}$ near the boundary at $\rho_{12}=0.$ }
\begin{align}
    \label{eq:GBBAdS}
    \mathcal{G}_{BB}^{AdS, \D}\left(\chi_1, \chi_2\right) = C_2\left(\D\right)\rho_{12}^{\D}\, \hypf\left(\D,\D-\frac{D-2}{2},2\D-D+2, \rho_{12}+i \varepsilon\right).
\end{align}
The normalization is fixed by demanding that the solution behaves like a $\delta$ function as $\chi_{12} \to 0$. This gives
\begin{align}
    \label{eq:C2def}
    C_2(\Delta) = \begin{cases}
        -\frac{i(-1)^D}{2\ell^{D-2}}\frac{\Gamma(\Delta)}{(-4)^\Delta\pi^{\frac{D-1}{2}} \Gamma(\Delta - \frac{D-3}{2})} \qquad \D \neq \frac{D-2}{2},\\
        -\frac{i(-1)^D}{4\ell^{D-2}}\frac{\Gamma(\Delta)}{(-4)^\Delta\pi^{\frac{D-1}{2}} \Gamma(\Delta - \frac{D-3}{2})} \qquad \D = \frac{D-2}{2}.
    \end{cases}
\end{align}
Note that unlike the flat space case, the bulk-to-bulk propagator depends on an additional parameter $\D$. It is easy to check that 
\begin{align}
    \label{eq:GBBAdSflatlim}
    \lim_{\ell \to \infty} \mathcal{G}_{BB}^{AdS, \D}\left(\chi_1, \chi_2\right) =  \mathcal{G}_{BB}^{Flat}\left(X_1, X_2\right),
\end{align}
independently of the value of $\D$. This has to be expected since $m \to 0$ in the limit we are considering.

\paragraph{Bulk-to-boundary propagator:}This is obtained from the bulk-to-bulk propagator \eqref{eq:GBBAdS} by sending one point to infinity and multiplying by appropriate powers of $r$ to keep it finite. 
\begin{align}
\label{eq:GBbAdS}
    \mathcal{G}^{AdS, \Delta}_{Bb,\epsilon}(\bar{X}_1, \chi_2) & \equiv  \ell^{\frac{D-2}{2}} \lim_{r_1 \to \epsilon\infty} \left(\frac{r_1}{\ell} \right)^{\Delta} \mathcal{G}^{AdS, \Delta}_{BB}(\chi_1, \chi_2) \\
    &\nonumber = \ell^{\frac{D-2}{2}} \, C_2(\Delta) \left( \frac{-4 \ell}{-\frac{1}{\ell^2}r_2 u_{12}^2 -2 u_{12} + r_2 \left|{\bf x}_{12}\right|^2+i \epsilon\varepsilon}\right)^\Delta.
\end{align}
Here $\epsilon=\pm 1$ and distinguishes incoming and outgoing particles. We can compute the flat limit of this propagator to get
\begin{align}
\label{eq:GBbAdSflatlimit}
    \lim_{\ell \to \infty} \mathcal{G}^{AdS, \Delta}_{Bb,\epsilon}(\bar{X}_1, \chi_2) =  \alpha_D\left(\D\right) \mathcal{G}_{Bb.\epsilon}^{Flat,\D}\left(x_1, X_2\right),
\end{align}
where 
\begin{align} \label{alpha value}
    \alpha_D\left(\D\right) =  i (-1)^{\D} \ell^{\D+\frac{D-2}{2}} 2^{1+\D+\frac{D}{2}} \frac{\pi^{\frac{D}{2}}}{\Gamma\left(\D\right)} C_2\left(\D\right) = \begin{cases}
        \frac{(-1)^{D} \sqrt{\pi} \ell^{\D-\frac{D-2}{2}}}{\Gamma\left(\D-\frac{D-3}{2}\right) 2^{\D-\frac{D}{2}}} \qquad &\D \neq \frac{D-2}{2},\\
         (-1)^{D} &\D = \frac{D-2}{2}.
    \end{cases}
\end{align}

In the next sections, we will use these propagators to define boundary correlators in AdS${}_D$ which we will denote by $\an{\mo^{\e_1}_{\D_1}\left(\bar{X}_1 \right) \dots \mo^{\e_n}_{\D_n}\left(X_n \right)}$. The representation of the propagators in Bondi coordinates is useful for computing the flat limit. However, in order to perform the integrals over the bulk points, it is necessary to Wick rotate to Euclidean AdS. The resulting correlators can be analytically continued to Lorentzian signature following well-known procedures \cite{Duffin}. We will denote Euclidean correlators by $\an{\mo_{\D_1}\left(\bar{X}_1 \right) \dots \mo_{\D_n}\left(\bar{X}_n \right)}_{E}$.

\subsection{Two-point correlator}
\label{sec:ads2ptcorr}

The $2$-point boundary correlator in AdS${}_D$ can be computed by sending the bulk point to infinity in the bulk-to-boundary propagator \eqref{eq:GBbAdS}:
\begin{equation}
\label{eq:2ptLorcorr}
    \langle O^{-1}_\Delta (\bar{X}_1) O^{1}_\Delta (\bar{X}_2) \rangle = \ell^{\frac{D-2}{2}}\lim_{r_2 \to \infty} \left( \frac{r_2}{\ell} \right)^\Delta\mathcal{G}^{AdS, \Delta}_{Bb, -}(\bar{X}_1, \chi_2) =  \frac{\tilde{C_2}(\Delta)}{(-\frac{1}{\ell^2} u_{12}^2 + \left|{\bf x}_{12}\right|^2 + i \varepsilon)^{\D}}
\end{equation}
where $\tilde{C_2}(\Delta) = (-4)^\Delta \ell^{D-2} C_2(\Delta)$, with $C_2\left(\D\right)$ defined in \eqref{eq:C2def}. Note that this is just a CFT${}_{D-1}$ $2$-point function with $\frac{1}{\ell}$ playing the role of the speed of light. The corresponding Euclidean correlator is 
\begin{equation}
\label{eq:2ptEuccorr}
    \langle O_\Delta (\bar{X}_1) O_\Delta (\bar{X}_2) \rangle_{E} = \ell^{\frac{D-2}{2}}\lim_{r_2 \to -\infty} \left( \frac{r_2}{\ell} \right)^\Delta\mathcal{G}^{AdS, \Delta}_{Bb, 1}({\bf x}_1, \chi_2) =  \frac{\tilde{C_2}(\Delta)}{(\frac{1}{\ell^2} t^2_{12} + \left|{\bf x}_{12}\right|^2)^{\D}} .
 \end{equation} The flat space limit of this will be discussed at the same time as the Carrollian limit in Section \ref{sec:Two-point functions lim}.

\subsection{Three-point correlator}
\label{sec:ads3ptcorr}

We can compute the 3-point correlator by integrating three bulk-to-boundary propagators over the common bulk point:
\begin{align}
     \label{eq:3ptLorcorr}
\an{\mo^{\e_1}_{\D_1}\left(\bar{X}_1\right)\mo^{\e_2}_{\D_2}\left(\bar{X}_2\right)\mo^{\e_3}_{\D_3}\left(\bar{X}_3\right)} &=\kappa_3 \beta_3 \int_{AdS_D} d^{D+1}\chi \, \delta\left(\chi^2+\ell^2\right) \prod_{i=1}^3 \mathcal{G}_{Bb,\e_i}^{AdS,\D_i}\left(\bar{X}_i, \chi\right)
\end{align}
However, the integration is best performed in Euclidean AdS and the resulting Euclidean CFT correlator is:
\begin{align}
    \label{eq:3ptEuccorr}
    \an{\mo_{\D_1}\left(x_1\right)\mo_{\D_2}\left(x_2\right)\mo_{\D_3}\left(x_3\right)}_{E} & =  \ell^{\frac{6-D}{2}} \kappa_3\, \mathcal{N}_{\D_1, \D_2, \D_3} \prod_{i<j}\left(\frac{t_{ij}^2}{\ell^2}+{\bf x}_{ij}^2\right)^{-\D_{ij}}\ ,
\end{align}
where  $\beta_3$ was defined in the paragraph below \eqref{eq:3ptfeyndiagdef} and 
\begin{align}
\label{eq:Nanddeltaijdef}
\nonumber &\mathcal{N}_{\D_1, \D_2, \D_3} = \beta_3\frac{\pi^{\frac{D-1}{2}}}{2}\Gamma\left(\frac{\Sigma_{\D}-(D-1)}{2}\right)\prod_{i<j}\Gamma\left(\D_{ij}\right)\prod_{k}\frac{\tilde{C}_2\left(\D_k\right)}{\Gamma\left(\D_k\right)},\\
&\qquad\qquad \D_{ij} = \D_i + \D_j - \frac{\Sigma_{\D}}{2}, \qquad \Sigma_{\D} = \sum_{i} \D_i.
\end{align}
Note that the scaling of $\ell^{\frac{6-D}{2}}$ in the definition of the three-point CFT correlator follows from the usual normalization of the AdS Witten diagrams, as in \cite{Penedones:2010ue}. In order the compute the flat limit from a bulk perspective, we first express the integral \eqref{eq:3ptLorcorr} in Bondi coordinates. There is a subtlety in taking the flat limit that we address here. This was also highlighted in \cite{Alday:2024yyj}. We are interested in computing the limit $\frac{r}{\ell} \to 0$, where $r$ corresponds to any length scale. In practice, we take this limit by first introducing a cutoff scale $\Lambda$ in the domain of integration, before taking the limit. We can then replace each bulk-to-boundary propagator by its flat limit using \eqref{eq:GBbAdSflatlimit}. If we now take $\Lambda \to \infty$, we get:  
\begin{align}
    \an{\mo^{\e_1}_{\D_1}\left(\bar{X}_1\right)\mo^{\e_2}_{\D_2}\left(\bar{X}_2\right)\mo^{\e_3}_{\D_3}\left(\bar{X}_3\right)} \xrightarrow[]{\ell \to \infty} &\kappa_3 \beta_3 \int_{Flat} d^{D}X \, \prod_{i=1}^3 \alpha_D\left(\D_i\right)\mathcal{G}_{Bb,\e_i}^{Flat, \D_i}\left(x_i, X\right) \\
    &\nonumber = \prod _{i=1}^3 \alpha_D\left(\D_i\right) \mathcal{C}_3^{\D_1, \D_2,\D_3}
\end{align}
which reproduces the integral representation of the $3$-point Carrollian amplitude \eqref{eq:3ptfeyndiagdef}. We will also demonstrate this result explicitly from the boundary CFT perspective in the next section.

\subsection{Four-point correlator}
\label{sec:ads4ptcorr}

The 4-point Lorentzian correlator corresponding to a contact interaction in $D$ dimensions is given by:
\begin{equation}
\label{eq:4ptLorcorr}
    \langle \mathcal{O}^{\e_1}_{\Delta_1}(\bar{X}_1) \mathcal{O}^{\e_2}_{\Delta_2}(\bar{X}_2) \mathcal{O}^{\e_3}_{\Delta_3}(\bar{X}_3) \mathcal{O}^{\e_4}_{\Delta_4}(\bar{X}_4) \rangle^c = \kappa_4 \beta_4 \int_{AdS_D} d^{D+1} \chi \, \delta\left(\chi^2+\ell^2\right) \prod_{i=1}^4 \mathcal{G}^{AdS,\Delta_i}_{Bb,\e_i} (\bar{X}_i, \chi)
\end{equation}
where $\beta_4$ has been defined earlier below \eqref{eq:4ptfeyndiagdef}. As before, the integration is best carried out in Euclidean AdS bulk with the result:
\begin{align}
\label{eq:4ptEuccorr}
    \langle \mathcal{O}_{\Delta_1}(\bar{X}_1) \mathcal{O}_{\Delta_2}(\bar{X}_2) \mathcal{O}_{\Delta_3}(\bar{X}_3) \mathcal{O}_{\Delta_4}(\bar{X}_4) \rangle^c_E &= \kappa_4 \,\ell^{4-D}\beta_4 \,  \mathcal{Z}\left(\bar{X}_{ij}\right)\mathcal{N}_4 \bar D_{\Delta_1, \Delta_2, \Delta_3, \Delta_4}(U,V) .
\end{align}
Similar to the 3-point case, the scaling of $\ell^{4-D}$ again follows from the usual normalization of 4-point contact AdS Witten diagrams. Here $U,V$ are the $D-1$ dimensional conformal cross ratios
\begin{align}
\label{eq:crossratios}
U = \frac{\left(\frac{u_{12}^2}{\ell^2}+{\bf x}_{12}^2\right)\left(\frac{u_{34}^2}{\ell^2}+{\bf x}_{34}^2\right)}{\left(\frac{u_{13}^2}{\ell^2}+{\bf x}_{13}^2\right)\left(\frac{u_{24}^2}{\ell^2}+{\bf x}_{24}^2\right)},\qquad V = \frac{\left(\frac{u_{14}^2}{\ell^2}+{\bf x}_{14}^2\right)\left(\frac{u_{23}^2}{\ell^2}+{\bf x}_{23}^2\right)}{\left(\frac{u_{13}^2}{\ell^2}+{\bf x}_{13}^2\right)\left(\frac{u_{24}^2}{\ell^2}+{\bf x}_{24}^2\right)} , 
\end{align}
$\mathcal{Z}\left(\bar{X}_{ij}\right)$ is a standard prefactor responsible for the correct transformation properties
\begin{equation}
\label{eq:confprefact}
    \mathcal{Z}\left(\bar{X}_{ij}\right) = \frac{\left(\frac{u_{14}^2}{\ell^2}+{\bf x}_{14}^2\right)^{\frac{1}{2}\Sigma_\Delta - \Delta_1 - \Delta_4} \left(\frac{u_{34}^2}{\ell^2}+{\bf x}_{34}^2\right)^{\frac{1}{2}\Sigma_\Delta - \Delta_3 - \Delta_4}}{\left(\frac{u_{13}^2}{\ell^2}+{\bf x}_{13}^2\right)^{\frac{1}{2}\Sigma_\Delta - \Delta_4} \left(\frac{u_{24}^2}{\ell^2}+{\bf x}_{24}^2\right)^{\Delta_2}},
\end{equation}
and the overall normalization is given by 
\begin{align}
\label{eq:normalization4ptAdS}
    \mathcal{N}_4 = \frac{\pi^{\frac{(D-1)}{2}}}{2} \Gamma\left( \frac{\Sigma_\Delta - (D-1)}{2}\right) \prod_{i=1}^4 \frac{\tilde{C_2}(\Delta_i)}{\Gamma(\Delta_i)}, 
\end{align}
with $\Sigma_\Delta = \sum_{i=1}^4 \Delta_i$. We can compute the flat limit from a bulk perspective as in the three-point case. This gives
\begin{align}
    \label{eq:4ptflatlimitbulk}
     \langle \mathcal{O}^{\e_1}_{\Delta_1}(\bar{X}_1) \mathcal{O}^{\e_2}_{\Delta_2}(\bar{X}_2) \mathcal{O}^{\e_3}_{\Delta_3}(\bar{X}_3) \mathcal{O}^{\e_4}_{\Delta_4}(\bar{x}_4) \rangle^c \xrightarrow[]{\ell \to \infty} &\kappa_4 \beta_4 \int d^{D+1}X \, \prod_{i=1}^4 \alpha_D\left(\D_i\right)\mathcal{G}_{Bb,\e_i}^{Flat, \D_i}\left(x_i, X\right) \\
    &\nonumber = \prod _{i=1}^4 \alpha_D\left(\D_i\right) \mathcal{C}_{4,c}^{\D_1, \D_2,\D_3,\D_4} 
\end{align} which reproduces the integral expression of the $4$-point Carrollian amplitude \eqref{eq:4ptfeyndiagdef}.

\section{Carrollian limit of CFT${}_{D-1}$ correlators}
\label{sec:carrlimitcftcorrs}

In this section, we will compute Carrollian limits of scalar correlators in a CFT in $D-1$ dimensions. We will show that the Carrollian limits of $2$-, $3$- and $4$-point correlators dual to bulk Witten diagrams (\eqref{eq:2ptEuccorr}, \eqref{eq:3ptEuccorr},\eqref{eq:4ptEuccorr}) reproduce the corresponding Carrollian amplitudes (\eqref{eq:2ptcarramp}, \eqref{eq:3ptcarramp}, \eqref{eq:4ptcarrramp}). Hence, combined with the results of the previous section, this demonstrates the correspondence between flat space limit in the bulk ($\ell \to \infty$) and Carrollian limit at the boundary ($c\to 0$) at the level of the correlators, extending the results of \cite{Alday:2024yyj} to general dimensions.

\subsection{Two-point function}
\label{sec:Two-point functions lim}

We can either directly start from the Lorentzian $2$-point function \eqref{eq:2ptLorcorr} (and replacing $\frac{1}{\ell}$ by $c$) or, alternatively, start from the Euclidean $2$-point function \eqref{eq:2ptEuccorr} and analytically continue to  Lorentzian signature to the time ordering in which operator 1 is in the past and 2 is in the future. This corresponds to the configuration in which operator 1 is incoming and 2 is outgoing. 
\begin{equation}
    \label{eq:to2ptfunc}
    \an{\mo^{-}_{\D_1}\left(\bar{X}_1\right)\mo^{+}_{\D_2}\left(\bar{X}_2\right)}  =  \frac{\tilde{C}_2\left(\D\right) \delta_{\D_1, \D_2}}{\left(-c^2 u_{12}^2 + \left|{\bf x}_{12}\right|^2+i\varepsilon \right)^{\D_1}} .
\end{equation}
Computing the Carrollian limit of this 2-point function is similar to the $D = 4$ case in \cite{Alday:2024yyj}. There are exactly two different possible Carrollian limits given by: 
\begin{align}
    \label{eq:mathstatements}
    \lim_{c \to 0} c^{\alpha} \an{\mo^{-}_{\D_1}\left(\bar{X}_1\right)\mo^{+}_{\D_2}\left(\bar{X}_2\right)} =  
    \begin{cases}
      \frac{  \tilde{C}_2\left(\D\right) \delta_{\D_1, \D_2}}{|{\bf x}_{12}|^{2\D_1} } \qquad &\alpha = 0 \ ,\\ 
      \frac{\pi^{\frac{D-2}{2}}\tilde{C}_2\left(\D\right) \Gamma\left(\D - \frac{D-2}{2}\right) }{\Gamma(\D)} \frac{\delta^{(D-2)}\left({\bf x}_{12}\right) }{\left(-u_{12}^2+i \varepsilon \right)^{\D-\frac{D-2}{2}}}  \qquad &\alpha = 2\D-D+2 \ .
    \end{cases} 
\end{align}
These are generalizations of the usual 2-point functions of Carrollian CFTs to $D-1$ dimensions. The first corresponds to the magnetic branch, and the second, the electric branch. Like the $D = 4$ case, one can note that the electric branch is only reached when the two operators are timelike separated. We can then formally define the Carrollian electric and magnetic operators $\Phi$ and $\Psi$ as
\begin{equation}
 \partial_u^{\D - \frac{D-2}{2}} \Phi^{\e}(x) \equiv \lim_{c \to 0}\frac{1}{ \alpha_D(\D) }\mo^{\e}_\D(\bar{X}) \ \ \ \ \ , \ \ \ \ \ \Psi^\epsilon(x) \equiv \lim_{c \to 0}\frac{1}{\sqrt{\tilde{C}_2(\D)}} \mo^\epsilon_\D (x).
\end{equation}
Notice the different scalings in $c$ for the electric and magnetic limits, where $\alpha (\Delta)$ is given in \eqref{alpha value}. With these definitions, we have: 
\begin{equation}
\begin{split}
    \langle \partial_{u_1}^{\D - \frac{D-2}{2}} \Phi^{-1}(x_1) \partial_{u_2}^{\D - \frac{D-2}{2}} \Phi^{+1}(x_2)\rangle &= \lim_{c \to 0} \frac{\an{\mo^{-1}_{\D}\left(x_1\right)\mo^{+1}_{\D}\left(x_2\right)}}{(\alpha_D(\D))^2} \\&= \frac{-i}{4\pi \kappa_2} \mathcal{C}_2^{\D,\D}
\end{split}
\end{equation}
which agrees with the $2$-point Carrollian amplitude \eqref{eq:2ptcarramp} if we set $\kappa_2 = \frac{i \pi}{2}$. 

\subsection{Three-point function}
\label{sec:carrllimit3ptfunc}

As with the two-point correlator, we must analytically continue the three-point correlator \eqref{eq:3ptEuccorr} to Lorentzian signature. Continuing to the configuration in which all three are timelike separated from each other, we get
\begin{align}
\label{eq:3ptCFTlor}
     \an{\mo^{\e_1}_{\D_1}\left(x_1\right)\mo^{\e_2}_{\D_2}\left(x_2\right)\mo^{\e_3}_{\D_3}\left(x_3\right)} & = \frac{\kappa_3}{c^{\frac{6-D}{2}}}\frac{\mathcal{N}_{\D_1, \D_2, \D_3}}{\displaystyle{\prod_{i<j}}\left(-c^2 u_{ij}^2+{\bf x}_{ij}^2+i \varepsilon \e_i \e_j\right)^{\D_{ij}} } ,
\end{align}
Recall that the scaling of $1/c^{\frac{6-D}{2}}$ follows from \eqref{eq:3ptEuccorr} with the identification $c \leftrightarrow 1/\ell$. The magnetic Carrollian limit is straightforward and gives 
\begin{align}
    \label{eq:3ptmagnetic}
    \lim_{c \to 0}  \an{\mo^{\epsilon_1}_{\D_1}\left(x_1\right)\mo^{\epsilon_2}_{\D_2}\left(x_2\right)\mo^{\epsilon_3}_{\D_3}\left(x_3\right)}  = \an{\Psi^{\epsilon_1}\left({\bf x}_1 \right)\Psi^{\epsilon_2}\left({\bf x}_2 \right)\Psi^{\epsilon_3}\left({\bf x}_3 \right)}& = \tilde{\kappa_3}\frac{\mathcal{N}_{\D_1, \D_2, \D_3}}{\displaystyle{\prod_{i<j}}\left({\bf x}_{ij}^2+i \varepsilon \e_i \e_j\right)^{\D_{ij}} },
\end{align}
where $ c^{\frac{6-D}{2}}\tilde{\kappa}_3 = \kappa_3$ is a rescaled coupling. We can have a non-trivial electric limit only on the support of singular configurations. In order to compute this electric limit, we will use a technique presented in Section [5] of \cite{deGioia:2024yne}. We start by rewriting the three-point function in a Schwinger parametrized form as: 
\begin{align}
    \label{eq:3ptfuncschwinger}
    \an{\mo^{\e_1}_{\D_1}\left(x_1\right)\mo^{\e_2}_{\D_2}\left(x_2\right)\mo^{\e_3}_{\D_3}\left(x_3\right)} &= \frac{\kappa_3 \mathcal{N}_{\D_1, \D_2, \D_3}}{c^{\frac{6-D}{2}}\prod_{i<j} \Gamma\left(\D_{ij}\right)} \int_{0}^\infty \prod_{j<k} d\om_{jk} \,\om_{jk}^{\D_{jk}-1} e^{i\om_{jk} \epsilon_j \epsilon_k \left( -c^2 u_{jk}^2+{\bf x}_{jk}^2\right) -\varepsilon \om_{jk}},\nonumber\\
    & = \frac{\kappa_3 \mathcal{N}_{\D_1, \D_2, \D_3}}{c^{\Sigma_{\D}+\frac{6-D}{2}}\prod_{i<j} \Gamma\left(\D_{ij}\right)}   \int_{0}^\infty \prod_{i} d\om_i \,\om_i^{\D_{i}-1} e^{i\sum_{j<k}\om_j \om_k \epsilon_j \epsilon_k \left( -u_{jk}^2+\frac{{\bf x}_{jk}^2}{c^2}\right) -\varepsilon \om_j \om_k}.
\end{align}
In the first line, we have introduced three Schwinger parameters $\om_{12}, \om_{23}, \om_{13}$ and in the second line, we changed variables to $\om_{jk} = \frac{\om_j \om_k}{c^2}$.
For the electric limit, we want to compute
\begin{align}
    \lim_{c \to 0}   \prod_i\left(\alpha_D\left(\D_i\right)\right)^{-1}\an{\mo^{\e_1}_{\D_1}\left(x_1\right)\mo^{\e_2}_{\D_2}\left(x_2\right)\mo^{\e_3}_{\D_3}\left(x_3\right)} &=   \frac{\kappa_3 \mathcal{N}_{\D_1, \D_2, \D_3}}{\prod_i \alpha_D\left(\D_i\right)}  \int_{0}^\infty \prod_{i} d\om_i \,\om_i^{\D_{i}-1}\\
    & \qquad\times \frac{1}{\prod_{i<j} \Gamma\left(\D_{ij}\right)}e^{-i\sum_{j<k}\om_j \om_k \epsilon_j \epsilon_k u_{jk}^2-\varepsilon \om_j \om_k}\mathcal{L},\nonumber
\end{align}
where
\begin{align}
\label{eq:ansatz3ptlimit}
   \mathcal{L} =  \lim_{c \to 0}\frac{1}{c^D}e^{i\sum_{j<k} \frac{\e_j \e_k \om_j \om_k}{c^2}{\bf x}_{jk}^2} =g(\om_i)  \left|{\bf x}_{12}\right|^{D-4} \delta^{D-2}\left({\bf x}_{12}\right)\delta^{D-2}\left({\bf x}_{13}\right).
\end{align}
The second equality is an ansatz motivated by the Carrollian amplitude \eqref{eq:3ptcarramp}. It is worth pausing here to comment on this ansatz and on the limit in general. Note that one can rewrite
\begin{equation}
    \frac{1}{c^2}\sum_{i<j} \om_i \om_j \epsilon_i \epsilon_j \Vec{\textbf{x}}_{ij}^2 = -\frac{1}{c^2} \left( \sum_{i = 1}^3 \epsilon_i \om_i q_i \right)^2 \equiv \frac{P^2}{c^2},
\end{equation}
where $P$ as defined above is total momentum. In the $c \to 0$ limit, we will pick up dominant contributions from the locus where $P$ is null, which does not imply $P = 0$. Indeed, if the $q_i$ are all collinear, then clearly $P^2 = 0$ but $P$ is in general non-zero. In \cite{deGioia:2024yne}, the authors focused on the term with $P=0$ which reproduced the correct Celestial amplitude. We will see the same holds true for the Carrollian case. However, in the limit, we expect more than just the Carrollian amplitude. We will see this expectation borne out explicitly below. We can evaluate $g\left(\om_i\right)$ by computing the following integral:
\begin{align}
    \label{eq:gint}
    g\left(\om_i\right) &= \int d^{D-2} {\bf x}_{12} \,d^{D-2} {\bf x}_{13} \,  \left|{\bf x}_{12}\right|^{4-D}  \frac{1}{c^D}e^{i\sum_{j<k} \frac{\e_j \e_k \om_j \om_k}{c^2}{\bf x}_{jk}^2}\\
    &= \frac{2\pi^{\frac{D-2}{2}}}{i^{\frac{D-4}{2}}\Gamma\left(\frac{D-2}{2}\right)}\frac{\left(\e_3\om_3\right)^{\frac{2-D}{2}}\left(\e_1\om_1+\e_2\om_2\right)^{\frac{4-D}{2}}}{\e_1\e_2\om_1 \om_2 \left(\e_1\om_1+\e_2\om_2+\e_3\om_3\right)} .
\end{align}
The integral over the $\om_i$ is now ill defined due to the pole at $\e_1\om_1+\e_2\om_2+\e_3\om_3=0.$\footnote{The branch point at $\e_1\om_1+\e_2\om_2=0$ also needs a prescription and will lead to a new contribution to the Carrollian limit. However, this contribution will not include the Carrollian amplitude and we will not discuss it in this work.} The result depends on the choice of  $i \varepsilon$ prescription. Indeed this can be explicitly demonstrated by using the well-known Sokhotsky formula: 
\begin{equation}
\label{eq:pvformula}
    \lim_{\varepsilon \to 0^+} \frac{1}{\epsilon_1 \om_1 + \epsilon_2 \om_2 + \epsilon_3 \om_3 \pm i\varepsilon} = \mp i \pi \delta(\epsilon_1 \om_1 + \epsilon_2 \om_2 + \epsilon_3 \om_3) + \text{p.v.} \frac{1}{\epsilon_1 \om_1 + \epsilon_2 \om_2 + \epsilon_3 \om_3} .
\end{equation}
We thus get
\begin{align}
\label{eq:carrlimit3ptcctilde}
     \lim_{c \to 0}   \prod_i\left(\alpha_D\left(\D_i\right)\right)^{-1}\an{\mo^{\e_1}_{\D_1}\left(x_1\right)\mo^{\e_2}_{\D_2}\left(x_2\right)\mo^{\e_3}_{\D_3}\left(x_3\right)} = \mathcal{C}_3^{\D_1, \D_2, \D_3} + \tilde{\mathcal{C}}_3.
\end{align}
$\mathcal{C}_3^{\D_1, \D_2, \D_3}$ is the Carrollian amplitude in \eqref{eq:3ptcarramp} which arises from the $\delta$ function piece in \eqref{eq:pvformula}. $\tilde{\mathcal{C}}_3$ is the contribution of the principal value piece. We present the details of the computation in  Appendix \ref{sec:ctildecalc}, merely reproducing the final result here
\begin{align}
    \tilde{\mathcal{C}}_3 = -&\frac{4\pi^2\tilde{\mathcal{K}}_3}{(-u_{23}^2)^{\frac{\Sigma_\Delta - D}{2}} }\delta^{D-2}\left({\bf x}_{12}\right)\delta^{D-2}\left({\bf x}_{13}\right) \left|{\bf x}_{12}\right|^{D-4} \times \\
    &\left[C_1  F_1 + C_2 \left(\frac{u_{13}}{u_{23}}\right)^{D-2-2\D_{13}} F_2  +  C_3 \left(\frac{u_{12}}{u_{23}}\right)^{2-2\D_{12}}  F_3   + C_4 \left(\frac{u_{13}}{u_{23}}\right)^{D-2-2\D_{13}} \left(\frac{u_{12}}{u_{23}}\right)^{2-2\D_{12}}  F_4\right] \nonumber,
\end{align}
where $C_i$ are constants defined in \eqref{eq:ctildeCi} and the $F_i$ are Kampe de Feriet functions whose explicit expressions can be found in \eqref{eq:ctildeF}. Note that the Carrollian amplitude $\tilde{C}_3$ depends only on one ratio $\frac{u_{32}}{u_{12}},$ while these functions depend on two. $\tilde{\mathcal{C}}_3$ is consistent with the Carrollian Ward identities derived in \cite{Bagchi:2023fbj, Nguyen:2023miw}. This can be seen from the form of $\tilde{C}_3$ as a series in $\frac{u_{13}}{u_{23}}$ and $\frac{u_{12}}{u_{23}}$ presented in \eqref{eq:ctildedoublesumrep}.   We will further comment on the unexpected extra contribution $\tilde{\mathcal{C}}_3$ while analyzing its celestial counterpart in Section \ref{sec:From Carrollian to celestial amplitudes} and in the Outlook section \ref{sec:Outlook}.

\subsection{Four-point function}
\label{sec:carrlimit4ptfunc}

The procedure to compute the Carrollian limit of the correlator \eqref{eq:4ptEuccorr} is the same as the one followed in \cite{Alday:2024yyj} in $D=4$. Recall that the Euclidean correlator takes the form: 
\begin{align}
    \langle \mathcal{O}_{\Delta_1}(\bar{X}_1) \mathcal{O}_{\Delta_2}(\bar{X}_2) \mathcal{O}_{\Delta_3}(\bar{X}_3) \mathcal{O}_{\Delta_4}(\bar{X}_4) \rangle^c_E &= \frac{\kappa_4}{c^{4-D}} \,\beta_4 \,  \mathcal{Z}\left(\bar{X}_{ij}\right)\mathcal{N}_4 \bar D_{\Delta_1, \Delta_2, \Delta_3, \Delta_4}(U,V) .
\end{align}
where $\beta_4, \mathcal{N}_4$ and $\mathcal{Z}(\bar{X}_{ij})$ are introduced below \eqref{eq:4ptEuccorr} and the scaling $1/c^{4-D}$ follows from the usual identification $c \leftrightarrow 1/\ell$. We first analytically continue the Euclidean correlator to Lorentzian signature by moving two of the operator positions, say $\bar{X}_1, \bar{X}_2$ to the future of the other two. This is expressed as a Lorentzian $\bar{D}_{\Delta_1, \Delta_2, \Delta_3, \Delta_4}$ function, whose leading singularity $\Phi^{ls}_{\Delta_1, \Delta_2, \Delta_3, \Delta_4}$ is given by the dimension agnostic formula
\begin{equation}
    \hat\Phi^{ls}_{\Delta_1, \Delta_2, \Delta_3, \Delta_4} = \mathcal{K}_\Delta \frac{Z^{\Delta_3 + \Delta_4 - 2} (1-Z)^{\Delta_1 + \Delta_4 - 2}}{(Z - \bar Z)^{\Sigma_\Delta - 3}}, \quad   \mathcal{K}_\Delta = \pi^{3/2} 2^{\Sigma_\Delta - 2} \Gamma\left(\frac{\Sigma_\Delta - 3}{2}\right) (-1)^{\Delta_1 + \Delta_3}.
\end{equation}
Here $Z, \Zb$ are defined in terms of the $D-1$ dimensional cross ratios \eqref{eq:crossratios} as 
\begin{equation}
\label{eq:ZZbdef}
U=Z \Zb, \qquad\qquad  V = (1-Z)(1-\Zb).    
\end{equation}
As in the $D=4$ case, the Carrollian limit is non-trivial only on special kinematic configurations. We can identify this configuration by noting that\footnote{We have dropped terms of $\mo\left(c^4\right)$ in \eqref{eq:Fcarrlimit} since they will not contribute to the Carrollian limit unless $\mathcal{F}$ also vanishes.}
\begin{align}
\label{eq:Fcarrlimit}
    \left(Z-\Zb\right)^2 = \left(z-\zb\right)^2 - 2c^2 \mathcal{F} + \mo\left(c^4\right),\qquad   \mathcal{F} = \left(z+\zb\right)\left(U_1-V_1\right)-2U_1,
\end{align}
with $z, \zb$ related to the cross ratios on the Celestial sphere defined in \eqref{eq:celspherecrossratios} and  
\begin{align}
    \frac{U_1}{U} = \frac{u_{12}^2}{\left|{\bf x}_{12}\right|^2}+\frac{u_{34}^2}{\left|{\bf x}_{34}\right|^2}-\frac{u_{13}^2}{\left|{\bf x}_{13}\right|^2}-\frac{u_{24}^2}{\left|{\bf x}_{24}\right|^2}, \qquad    \frac{V_1}{V} = \frac{u_{23}^2}{\left|{\bf x}_{23}\right|^2}+\frac{u_{14}^2}{\left|{\bf x}_{14}\right|^2}-\frac{u_{13}^2}{\left|{\bf x}_{13}\right|^2}-\frac{u_{24}^2}{\left|{\bf x}_{24}\right|^2}.
\end{align}
We can have non-trivial behavior as $c \to 0$ on the support of the locus $\left(z-\zb\right)^2 = 0$. While this might seem like one constraint, it necessarily imposes $D-3$ constraints on the points ${\bf x}_1, \dots , {\bf x}_4$ of the celestial sphere. To see this, consider the Gram matrix $G$ defined by $G_{ij} = \left|{\bf x}_i -{\bf x}_j\right|^2 = -2q_i \cdot q_j$. We have
\begin{align}
\label{eq:gramdet}
\det G = \frac{\left(z-\zb\right)^2}{\left|{\bf x}_{13}\right|^4\left|{\bf x}_{24}\right|^4}  =  \frac{\left(z-\zb\right)^2}{16 \left(q_1\cdot q_3\right) \left(q_2 \cdot q_4\right)}.
\end{align}
This is a positive-definite matrix and its determinant vanishes if and only if the vectors $q_1, \dots , q_4$ are linearly dependent. The vectors $q_i$ are the embedding space vectors corresponding to the points ${\bf x}_i$. To be more precise, linear dependence requires the $D$ conditions $q_4^{\mu} = \sum_{i=1}^3 c_i q_i^{\mu}$which imposes $D-3$ constraints on $q_4$. In other words, the determinant of the Gram matrix $G$ being zero implies that there exist $c_i \in \mathbb{R}$ such that: 
\begin{equation}
    \sum_{i = 1}^4 c_i q_i^\mu = 0 \ \ \ \ \ \ \ \mu = 1,2\ldots, D
\end{equation}
In fact, this implies that the singular locus $(z - \zb)^2 = 0$ is equivalent to $\sum_{i = 1}^4 c_i q_i^\mu = 0$ for arbitrary $c_i \in \mathbb{R}$ satisfying this constraint. Hence, this leads to the ansatz: 
\begin{equation}
    \lim_{c \to 0} c^\alpha \hat\Phi^{ls}_{\Delta_1, \Delta_2, \Delta_3, \Delta_4} = \int \prod_{j = 1}^4 d c_j \delta^{(D)}\left(\sum_{i = 1}^4 c_i q_i^\mu \right) \mathcal{R}_D(c_i, q_i)
\end{equation}
where $\mathcal{R}_D(c_i, q_i)$ is some particular function of $c_i, q_i$, which we are denoting here just for schematics; and $\alpha$ is an appropriate scaling of $c$ that makes the right-hand side finite and non-zero. Note that with $c_i = \epsilon_i \om_i$ (with $\epsilon_i = \pm 1$ and $\om_i \in \mathbb{R}_+$, this is precisely the structure that a $D$-dimensional 4-point Carrollian amplitude has! Recall that from the point of view of the bulk, the $q_i$ are related to the momenta of the four bulk scalars as $p_i = \e_i \om_i q_i$, and the vanishing of the determinant is precisely telling us that the momenta of the four bulk scalars are linearly dependent. This is the origin of the $\delta$ functions in \eqref{eq:4ptdeltafunction}. Ofcourse, since there are more constraints imposed by the delta functions than there are $c_i$, one can exactly solve for the $c_i$ in terms of the variables $x_i$, barring an overall scale factor which needs to be integrated out separately. \\
\\
The above discussion clearly shows that the bulk-momentum conserving $\delta$ functions can be recovered from the Carrollian limit of the 4-point CFT correlator. However, we want to precisely derive the full Carrollian amplitude. This is most easilt done by choosing a special conformal frame for the four points on the boundary. To this end, first note that the constraint imposed by the delta functions also implies that: 
\begin{equation}
    \sum_{i = 1}^3 c_i ({\bf x}_i  - {\bf x}_4) = 0
\end{equation}
which is precisely the statement that the four points lie on a two-dimensional subspace of the celestial sphere $S^{D-2}$. Without loss of generality, we can consider the four points to be located at
\begin{align}
    \label{eq:qsinbasis}
    &{\bf x}_i = \left(x_i^1, x_i^2, 0, \dots , 0\right), \qquad i = 1, 2, 3,\\
    &\nonumber {\bf x}_4 = \left(x_4^1, x_4^2, 0, \dots , 0\right) +  \left(0, 0, x_4^3, \dots , x_4^{D-2}\right) \equiv {\bf x}_{4,s} + {\bf x}_{4,n}.
\end{align}
The subscript $s$ indicates the scattering plane, i.e. the plane formed by the points ${\bf x}_1, {\bf x}_2 , {\bf x}_3$. We can define cross-ratios in the scattering plane by
\begin{align}
    z_s\zb_s = \frac{\left|{\bf x}_{12}\right|^2\left|{\bf x}_{34,s}\right|^2}{\left|{\bf x}_{13}\right|^2\left|{\bf x}_{24,s}\right|^2}, \qquad (1-z_s)(1-\zb_s) = \frac{\left|{\bf x}_{12}\right|^2\left|{\bf x}_{34,s}\right|^2}{\left|{\bf x}_{13}\right|^2\left|{\bf x}_{24,s}\right|^2}.
\end{align}
Note that this choice is equivalent to choosing the vectors $n_j$ (See \eqref{eq:constraints}) to be
\begin{align}
\label{eq:njchoice}
    n_j^{\mu} = \delta^{\mu}_{j-2}, \qquad j=5, \dots , D.
\end{align}
With this explicit choice of conformal frame, the $\delta$ functions in the Carrollian amplitude \eqref{eq:4ptdeltafunction} become 
\begin{align}
    \delta\left(z-\zb\right)\prod_{j=5}^D \delta\left(q_4 \cdot n_j\right) =  \delta\left(z_s-\zb_s\right) \delta^{D-4}\left({\bf x}_{4,n}\right).
\end{align}
As explained earlier, these are also the constraints imposed by linear dependence of $q_1, \dots , q_4$. This leads us to the following claim that there exists a value of $\alpha$ for which 
\begin{align}
    \lim_{c \to 0} c^{\alpha}  \hat\Phi^{ls}_{\Delta_1, \Delta_2, \Delta_3, \Delta_4} = \mathcal{R}\left(u_{ij}, {\bf x}_{ij}\right)\,  \delta\left(z_s-\zb_s\right) \delta^{D-4}\left({\bf x}_{4,n}\right).
\end{align}
The function $\mathcal{R}$ and the exponent $\alpha$ can now be explicitly determined by performing the integral 
\begin{align}
    \mathcal{R} =c^{\alpha} \int dz_s \, d^{D-4} {\bf x}_{4,n}\,\hat\Phi^{ls}_{\Delta_1, \Delta_2, \Delta_3, \Delta_4} =  \mathcal{K}_\Delta Z^{\Delta_3 + \Delta_4 - 2} (1-Z)^{\Delta_1 + \Delta_4 - 2} \tilde{\mathcal{R}}
\end{align}
where
\begin{align}
   \tilde{\mathcal{R}} &= c^{\alpha}\int dz_s \, d^{D-4} {\bf x}_{4,n}\,\frac{1}{\left((z-\zb)^2+2c^2 \mathcal{F}\right)^{\frac{\Sigma_{\D}-3}{2}}} .
\end{align}
In order to evaluate this integral, we must work out how $\left(z-\zb\right)^2$ depends on $z_s, \zb_s.$ This is given by 
\begin{align}
    \left(z-\zb\right)^2 = \left(z_s-\zb_s\right)^2 - \sigma_1 \left|{\bf x}_{4,n}\right|^2 + \sigma_2 \left|{\bf x}_{4,n}\right|^4,
\end{align}
where $\sigma_1, \sigma_2$ are functions of $u_{ij}, {\bf x}_{ij,s}$. It is sufficient to determine $\sigma_1$ only (see \eqref{eq:rhoandF} for the explicit expression), since we can redefine the integration variables to $c w=z_s-\zb_s$ and  $c\tilde{{\bf x}}_{4,n} = {\bf x}_{4,n}$, in which case the contribution from $\sigma_2 \left|{\bf x}_{4,n}\right|^4$ becomes subleading in $c$. This simplifies the integral to give
\begin{align}\label{Rtilde4pt}
      \tilde{\mathcal{R}} &= c^{\alpha+D-\Sigma_{\D}}\int dw \, d^{D-4} \tilde{{\bf x}}_{4,n}\,\frac{1}{\left(w^2-\sigma_1\left|{\bf x}_{4,n} \right|^2+2\mathcal{F}\right)^{\frac{\Sigma_{\D}-3}{2}}}\nonumber \\
   &= c^{\alpha+D-\Sigma_{\D}}2^{4-\Sigma_{\D}} \pi^{\frac{D-3}{2}} \frac{\Gamma\left(\frac{\Sigma_{\D}-D}{2}\right)}{\Gamma\left(\frac{\Sigma_{\D}-3}{2}\right)}\frac{ \left(\frac{\left|z_s\right|^2 \left|{\bf x}_{23}\right|^2}{\left|{\bf x}_{34,s}\right|^2\left|{\bf x}_{24,s}\right|^2}\right)^{\frac{4-\Sigma_{\D}}{2}}}{\mathcal{U}^{\Sigma_{\D}-D}},
\end{align}
where we have used:
\begin{align}
    \label{eq:rhoandF}
    \sigma_1= 4\frac{\left|z_s\right|^2 \left|{\bf x}_{23}\right|^2}{\left|{\bf x}_{34,s}\right|^2\left|{\bf x}_{24,s}\right|^2}, \qquad \mathcal{F}\left|_{z=\zb} \right. = 2\frac{\left|z_s\right|^2 \left|{\bf x}_{23}\right|^2}{\left|{\bf x}_{34,s}\right|^2\left|{\bf x}_{24,s}\right|^2} \mathcal{U}^2,
\end{align}
with $\mathcal{U}$ defined in \eqref{eq:Udef}. We see that we must set $\alpha = \Sigma_{\D}-D$ to have a finite and non-zero limit. Putting all of this together,  we finally get 
\begin{align}
    \lim_{c \to 0} \frac{\an{\mo_{\D_1}^{\e_1}\left(x_1\right) \dots \mo_{\D_4}^{\e_4}\left(x_4\right) }}{\prod_{i=1}^4 \alpha_D\left(\D_i\right) }&= \lim_{c \to 0} \frac{\kappa_4}{c^{4-D}} \frac{c^{D-\Sigma_\D}}{\prod_{i=1}^4 \alpha_D\left(\D_i\right) }\,\beta_4 \,  \mathcal{Z}\left(\bar{X}_{ij}\right)\mathcal{N}_4 \mathcal{K}_\Delta z^{\Delta_3 + \Delta_4 - 2} (1-z)^{\Delta_1 + \Delta_4 - 2} \tilde{\mathcal{R}} \\&= \mathcal{C}_4^{\D_1, \dots \D_4}.
\end{align} 
where, in going to the last equality we have noted that $\prod_{i=1}^4 \alpha_D(\D_i)$ also has a $c$-scaling of $c^{2(D-2) - \Sigma_\D}$, which precisely cancels the $c-$scaling, hence yielding a finite and non-zero answer in the $c \to 0$ limit. Therefore, after an appropriate analytic continuation of \eqref{eq:4ptEuccorr} to Lorentzian signature, the electric Carrollian limit exactly reproduces the Carrollian $4$-point amplitude \eqref{Carrollian 4pt function}. It is worth stressing that the explicit formula of the Carrollian limit of the 4-point CFT$_{D-1}$ correlator derived here has wide applicability. One can now apply this formula to a variety of interesting 4-point holographic CFT$_{D-1}$ correlators to derive features of the general top-down Carrollian holograms from AdS$_D$/CFT$_{D-1}$ at the level of correlation functions. 

\section{From Carrollian to celestial amplitudes}
\label{sec:From Carrollian to celestial amplitudes}

Up to this stage, we have been focusing on Carrollian holography and showed that this framework is naturally related to AdS/CFT via a limiting procedure. We now relate the Carrollian boundary operators and amplitudes defined in Sections \ref{sec:carrboundops} and \ref{sec:carramps} to celestial operators and amplitudes in general dimensions. This relationship generalizes the one in $D=4$ presented in \cite{Donnay:2022wvx, Bagchi:2022emh, Donnay:2022aba}.

\subsection{Celestial operators}

For a massless scattering, boundary values of bulk fields, identified with Carrollian primaries, are in one-to-one correspondence with celestial primaries, as they are related via an invertible integral transform, which is the combination of a Fourier and a Mellin transform \cite{Donnay:2022wvx, Donnay:2022aba}. We have seen above that $u-$descendants of Carrollian primaries are also primaries (see below \eqref{carrollian primary}), which allows to generate more Carrollian correlators as in \eqref{descendant Carrollian amplitude}. Although the latter do not contain extra information on the massless scattering amplitudes, they are useful if one analytically continues $\Delta$ to complex values. In that case, the celestial amplitudes can just be found by setting $u=0$ \cite{Banerjee:2018gce,Banerjee:2018fgd, Bagchi:2022emh}. As we have computed these extra Carrollian correlators in Section \ref{sec:carramps}, we can simply apply this procedure here.

We start by recalling the expression for boundary operators at $\scri^{\pm}$ defined in \eqref{eq:boundaryopdef},
\begin{align}
     &\Phi^{\e}(u,{\bf x})  =  \int_0^\infty \frac{d \om}{2\pi} \,\left(-i \e\om\right)^{\frac{D-4}{2}} a^{\e}\, (\om, {\bf x}) e^{-i\e \om u -\varepsilon \om},
\end{align}
where $\e = \pm 1$ and $a^{+1} = a$ and $a^{-1} = a^{\dagger}$. We will use these to define the family of operators
\begin{align}
\label{eq:celestialprimarydef}
  \mbo_{\D}^{\e}\left({\bf x}\right) \equiv 2\pi\left.\partial_u^{\D-\frac{D-2}{2}}  \Phi^{\epsilon}(u,{\bf x}) \right|_{u=0} =  \int_0^\infty d\om\left(-i \e\om\right)^{\D-1}e^{-\varepsilon \om}a^{\e}(\om, {\bf x}) .
\end{align}
For $\D \in \mathbb{C}$, the derivative in the above equation should be understood to be defined via the integral. The integral transform in the right-hand side is just a Mellin transform. These operators $\mbo_{\D}^{\e}\left({\bf x}\right)$ transform as $SO(D-2)$ conformal primaries. To see that this is indeed the case, consider a bulk Lorentz transformation under which the coordinates at $\scri^\pm$ transform as  
\begin{align}
    \label{eq:bulklorentz}
    u \to u' = \left| \frac{\partial {\bf x}'}{\partial{\bf x}}\right|^{\frac{1}{D-2}}, \qquad {\bf x} \to {\bf x}'.
\end{align}
An argument along the lines of \eqref{eq:boundaryoptransform1}-\eqref{eq:boundaryoptransform2} shows that under such a transformation, we have 
\begin{align}
     \mbo_{\D}^{\e}\left({\bf x}\right) \to  \mbo_{\D}^{'\e}\left({\bf x}'\right)\equiv 2\pi \partial_{u'}^m \Phi^{'\epsilon}(u',{\bf x})'\left. \right|_{u=0}  = \left|\frac{\partial {\bf x}'}{\partial {\bf x}} \right|^{-\frac{\D}{D-2}} \mbo_{\D}^{\e}\left({\bf x}\right). 
\end{align}
Since Lorentz transformations in the bulk correspond to conformal transformations on the celestial sphere $S^{D-2},$ the above equation shows that the operators $\mbo_{\D}\left({\bf x}\right)$ indeed transform as conformal primaries with conformal weight $\D$. Under a bulk translation $ u \to u' = u -\xi$ with $\xi=q^{\mu}  t_{\mu}$, we have \cite{Donnay:2022wvx} 
\begin{align}
   \mbo_{\D}^{\e}\left({\bf x}\right)  \equiv 2\pi \partial_u^{\D - \frac{D-2}{2}} \Phi^{\e}\left(u, {\bf x}\right) \left. \right|_{u=0} \to  2\pi \partial_u^{\D - \frac{D-2}{2}}\Phi^{\e}\left(u, {\bf x}\right) \left. \right|_{u=\xi} = e^{i \e \partial_{\D}\xi }  \mbo_{\D}^{\e}\left({\bf x}\right)  .
\end{align}
This is the expected transformation law of conformal primaries in the celestial CFT \cite{Stieberger:2018onx}.

\subsection{From Carrollian to celestial amplitudes}

Using the celestial primary operators defined in \eqref{eq:celestialprimarydef}, we can define the celestial amplitude
\begin{align}
\label{eq:celampdef}
    \mathcal{M}_n^{\D_1, \dots , \D_n} &\equiv {}_{\text{out}}\an{0 \left|\mbo^{\epsilon_1}_{\D_1}\left({\bf x}_1\right) \dots \mbo^{\epsilon_n}_{\D_n}\left({\bf x}_n\right)\right|0}_{\text{in}} =  \prod_{i=1}^n \int d\om_i \left(-i\e_i\om_i\right)^{\D_i-1} \mA_n.
\end{align}
Here $\mA_n$ is the momentum space amplitude and the last equality follows from reasoning identical to that explained at the beginning of Section \ref{sec:carramps}. From the above discussion, the celestial amplitude transforms as a correlator of primaries in the celestial CFT. Comparing this definition with the correlator of Carrollian primaries with $u-$descendants \eqref{descendant Carrollian amplitude}, we arrive at the following simple relationship between Carrollian and celestial amplitudes \cite{Banerjee:2018gce,Banerjee:2018fgd},
\begin{align}
\label{eq:carrcelrelation}
      \mathcal{M}_n^{\D_1, \dots , \D_n} = \lim_{u_i \to 0} \left(2\pi\right)^n \mathcal{C}_n^{\D_1, \dots \D_n} .
\end{align}
It is important to consider a limiting procedure to obtain celestial amplitudes from Carrollian amplitudes, because the Carrollian amplitudes, when evaluated exactly at $u_i = 0$, are often ill-defined. 

\subsection{Two-, three- and four-point celestial amplitudes}

We can now use the relationship in \eqref{eq:carrcelrelation} to write down explicit expressions for celestial amplitudes directly from the Carrollian ones in \eqref{eq:2ptcarramp}, \eqref{eq:3ptcarramp}, \eqref{eq:4ptcarrramp}.

\paragraph{Two-point function:} The two-point amplitude is:
\begin{align}
\label{eq:2ptcelamp}
    \mathcal{M}_2^{\D_1, \D_2} &=2\kappa_2 i^{D} (-1)^{\D_1}\delta^{D-2}\left({\bf x}_{12}\right) \delta_{\e_1, -\e_2} \times \left[ \lim_{\varepsilon, u_i \to 0}\frac{\Gamma\left(\Sigma_{\D}-(D-2)\right)}{\left(u_{12}-i \e_1 \varepsilon\right)^{\Sigma_{\D}-(D-2)}}\right] \\
    &\nonumber =4\pi \kappa_2 i^{D+1} (-1)^{\D_1+1}\delta^{D-2}\left({\bf x}_{12}\right) \delta\left(\Sigma_{\D}-\left(D-2\right)\right)\delta_{\e_1, -\e_2}.
\end{align}
In arriving at this formula, we have made use of the distributional identity $\lim_{\nu \to 0} \Gamma\left(x\right)\nu^{-x} = -2\pi i \delta\left(x\right).$ The formula  \eqref{eq:2ptcelamp} is in agreement with (3.16) of \cite{deGioia:2024yne} upto an overall normalization.

\paragraph{Three-point function:}  Similarly the three-point celestial amplitude can also be obtained from the the Carrollian one \eqref{eq:3ptcarramp}. It is important to recall that in writing down the 3-point Carrollian amplitude in terms of the hypergeometric function, we required a specific ordering: $u_{32} < u_{12}$, which must also be respected as we take $u_i \to 0$ to recover the celestial amplitude. Hence, in this limiting procedure, we first take $u_{32} \to 0$, and then take $u_{12} \to 0$. Keeping this in mind and noting that
\begin{align}
    \lim_{z \to 0} \hypf(a,b,c;z) = 1 \ \ \ \text{and} \ \ \ \lim_{\nu\to 0} \Gamma(x) \nu^{-x} = -2\pi i \delta(x) 
\end{align}
we get the final form of the three-point celestial amplitude as
\begin{align}
\label{eq:3ptcelamp}
      \mathcal{M}_3^{\D_1, \D_2, \D_3} &= 2\kappa_3 \pi^{\frac{D}{2}}\left(-i\right)^{D-2}\,\frac{ B(\D_2 - 1, \D_3 - D+3)}{\Gamma\left(\frac{D-2}{2}\right)}  \left|{\bf x}_{12}\right|^{D-4}  \nonumber\\
    &\qquad \times \delta^{D-2}\left({\bf x}_{12}\right) \delta^{D-2}\left({\bf x}_{23}\right) \delta\left(\Sigma_\D - D\right).
\end{align}
This matches with the $D$-dimensional three-point celestial amplitude in \cite{deGioia:2024yne} up to a normalization. We can also compute $\tilde{\mathcal{M}}_3$, the Celestial counterpart of $\tilde{C}_3$ appearing in \eqref{eq:carrlimit3ptcctilde}. First, note that: 
\begin{align}
    \lim_{x,y \to 0} F^{2;0,1}_{0;1,2}(\Vec{\textbf{a}_{(p)}}, \Vec{\textbf{b}_{(q)}}; \Vec{\textbf{c}_{(r)}}, \Vec{\textbf{d}_{(s)}}; \Vec{\textbf{c}'_{(k)}}, \Vec{\textbf{d}'_{(l)}}; x, y) = 1 .
\end{align}
Consider the ordering $u_{2} < u_{1} < u_3$, which guarantees that $|u_{12}|,|u_{13}| < |u_{23}|$. We now take $u_{12}, u_{13} \to 0$ first, and then take $u_{23} \to 0$, which guarantees that the limiting procedure $u_i \to 0$ respects the ordering. Using this, we can write down the celestial counterparts of $\tilde{C}_3$ to be

\begin{align} 
\label{finalLorentzCelestial}
    \tilde{\mathcal{M}}_3 &= -4 \pi i^{D - \Sigma_\D + 1}\tilde{\mathcal{K}}_3 \delta^{D-2}\left({\bf x}_{12}\right)\delta^{D-2}\left({\bf x}_{13}\right) \left|{\bf x}_{12}\right|^{D-4} \\
    &\left[ \pi \csc\pi \D_{12} B\left(\D_2-1,\D_1-1\right)\delta\left(\Sigma_\Delta - D \right)  \right.  \nonumber \\  
    &\,\,\left. -2\pi^2 i \delta\left(\D_1+\D_3-(D-1)\right) \delta(\D_2 - 1) \csc\frac{\pi}{2}(D-2\D_1) , \nonumber \right. \\ 
    &\,\,\nonumber \left.-2i \pi \delta\left(\D_1+\D_2-\frac{D-2}{2}\right) \delta\left(\D_3-\frac{D-2}{2}\right) B\left(\D_1-1,\frac{D}{2}-\D_1\right), \nonumber  \right. \\
    \nonumber & \left. \hspace{0.2cm}  - 16\pi \delta\left(\D_{1}-\frac{D}{2}\right) \delta(\D_{2}-1) \delta \left(\D_{3}-\frac{D-2}{2}\right)\right] ,
\end{align}
where $\tilde{\mathcal{K}}_3 = \frac{\mathcal{K}_3 \pi^{D-2}}{\Gamma\left(\D_{23}\right)\Gamma\left(\frac{D}{2}-1\right) }$ with $\mathcal{K}_3$ defined in \eqref{kappa3}. As in the Carrollian case, it is unclear how to interpret these extra terms. The presence of multiple $\delta$ functions in the conformal dimensions indicates that this might be related to the soft sector of the 3-point amplitude, along the lines of \cite{Chang:2022seh}. This expression is not the Mellin transform of a momentum space amplitude. Consequently, it isn't clear how one would check if it was invariant under translations. We leave a complete understanding of these extra terms for future work. 
\paragraph{Four-point function:} Using the same logic, we can easily write down the four-point celestial amplitude by taking the $u_i \to 0$ limit of the four-point Carrollian amplitude \eqref{Carrollian 4pt function}: 
\begin{align}
    \label{eq:4ptcelamp}
    \mathcal{M}_4^{\D_1, \D_2, \D_3, \D_4} &=  \frac{-i\pi\kappa_4}{4}  \mathcal{S}\,\mathcal{Z}\left({\bf x}_{ij}\right)\delta\left(z-\zb\right)\prod_{j=5}^D \delta\left(q_4 \cdot n_j\right) \\ 
  &\qquad\times \left(-1\right)^{\D_1+\D_3-D} z^{\D_1-\D_2} \left(1-z\right)^{\D_2-\D_3} \times \left[\lim_{u_i \to 0}\frac{\Gamma\left(\Sigma_{\D}-D\right)}{\mathcal{U}^{\Sigma_{\D}-D}}\right]\nonumber\\
  &\nonumber = \frac{-\pi^2\kappa_4}{2} \mathcal{S}\,\mathcal{Z}\left({\bf x}_{ij}\right)\delta\left(z-\zb\right)\prod_{j=5}^D \delta\left(q_4 \cdot n_j\right) \\ 
  &\qquad\times \left(-1\right)^{\D_1+\D_3-D} z^{\D_1-\D_2} \left(1-z\right)^{\D_2-\D_3} \times \delta\left(\Sigma_{\D}-D\right) . \nonumber
\end{align} 
To the best of our knowledge, this expression, valid in any dimension, has not appeared earlier in the literature.

\section{Outlook}
\label{sec:Outlook}

In this paper, we have presented the Carrollian holography dictionary for massless scalar fields in $D \geq 4$ dimensions and defined the notion of Carrollian amplitudes, extending the results of \cite{Donnay:2022wvx, Mason:2023mti}. We also derived explicit expressions for celestial amplitudes, extending the relation between Carrollian and celestial holography established in \cite{Donnay:2022aba, Bagchi:2022emh, Donnay:2022wvx}. Furthermore, we have shown, in general dimensions, how Carrollian holography emerges naturally from AdS/CFT through a correspondence between the flat space limit in the bulk and a Carrollian limit at the boundary, extending the analysis of \cite{Alday:2024yyj}. In $D=4$, this correspondence allowed us to derive features of the top-down Carrollian hologram \cite{Lipstein:2025jfj} from AdS$_4$/CFT${_3}$ by considering the Carrollian limit of ABJM correlators. Our present analysis offers a clear setup to push this further and consider the flat space/Carrollian limit of AdS/CFT in other dimensions. More concretely, \eqref{eq:mathstatements}, \eqref{eq:carrlimit3ptcctilde} and \eqref{Rtilde4pt} -- which provide explicit expressions for the Carrollian limit of massless scalar holographic CFT correlators, can be leveraged to compute the Carrollian limit of any known holographic CFT$_{D-1}$, at least at the level of two, three, and four-point correlators. Notably, the $D=5$ case is of major interest as it constitutes the foundational example of holography \cite{Maldacena:1997re}, and many results are available on the CFT side for $\mathcal{N}=4$ Super-Yang-Mills theory.

The current set-up does not address the important question of the fate of the compactified dimensions that appear in explicit holographic dualities. For instance, in $D=4$, the bulk theory is AdS$_4 \times S^7$, and in $D=5$, it is AdS$_5 \times S^5$. In these cases, the radius of the sphere is related to the radius of AdS. Therefore, taking the flat space limit leads to a decompactification of the sphere, and we expect to land on a higher-dimensional flat space in the limit, with the KK modes yielding an infinite tower of massless fields. This technical issue was discussed in \cite{Lipstein:2025jfj}: the Carrollian limit captures scattering amplitudes restricted to the hyperplane coming from the limit of AdS. The transverse directions come from the decompactification of the sphere. We hope to come back to this question in the future.

A curious result that we found is that the Carrollian limit of the 3-point function does not match the 3-point Carrollian amplitude, see \eqref{eq:carrlimit3ptcctilde}. The two expressions differ by $\tilde{\mathcal{C}}_3$, which is explicitly written in Appendix \ref{sec:ctildecalc}. Notice that this extra contribution is compatible with the freedom left in the Ward identities to fix the electric 3-point function, see e.g. \cite{Nguyen:2023miw, Bagchi:2023fbj}. A further interesting observation is that the computation of $\tilde{C}$ is strikingly similar to the one encountered in computing correlation functions in non-conformal D$_p$-brane holography \cite{Bobev:2025idz}. A notable point of difference worth mentioning is that correlation functions in the strong-coupling limit in these non-conformal theories are in general obtained by integrating CFT correlators over auxiliary spacetime dimensions, whereas in the computation of $\tilde{C}$ (See \eqref{GuiLorentz}), the integration is performed over physical spacetime dimensions. On the other hand, the two-point function on the electric branch of Carrollian CFTs (See \eqref{eq:mathstatements}) does precisely take the form of two-point functions in non-conformal field theories having a generalized conformal structure. Whether this is a mere coincidence or hints towards some kind of generalized conformal structure \cite{Jevicki:1998ub, Taylor:2017dly} of Carrollian CFTs is worth exploring. In addition, it would be interesting to understand the precise role of $\tilde{\mathcal{C}}_3$ and the implications for scattering processes in flat space.

\section*{Acknowledgments} 
It is our pleasure to thank Luis Fernando Alday, Burkhard Eden, Paul Heslop, Arthur Lipstein and Ana-Maria Raclariu for discussions and/or collaborations on related topics. The work of H.K. is supported by the Clarendon Fund Scholarship and the Eddie-Dinshaw Scholarship at Balliol College. R.R. is supported by the Titchmarsh Research Fellowship at the Mathematical Institute and by the
Walker Early Career Fellowship at Balliol College.

\appendix
\section{Computation of $\tilde{\mathcal{C}}_3$}
\label{sec:ctildecalc}
In this Appendix, we compute the contribution $\tilde{\mathcal{C}}_3$ in \eqref{eq:carrlimit3ptcctilde}. Instead of computing the Schwinger parameterized integral, we will find it convenient to start from the position space representation \eqref{eq:3ptLorcorr} and compute the integrals over ${\bf x}_i$ directly. We make the ansatz:
\begin{align}
    \label{eq:3ptposspace}
\lim_{c \to 0} \frac{\an{\mo_{\D_1}\left(x_1\right)\mo_{\D_2}\left(x_2\right)\mo_{\D_3}\left(x_3\right)}}{\prod_{i=1}^3\alpha_D\left(\D_i\right)} =\mathcal{G}\left(u_i\right)\delta^{D-2}\left({\bf x}_{12}\right)\delta^{D-2}\left({\bf x}_{13}\right) \left|{\bf x}_{12}\right|^{D-4}.
\end{align}
For more explanation about the ansatz, see \eqref{eq:ansatz3ptlimit}. We can now compute the function $\mathcal{G}\left(u_i\right)$ by integrating the CFT 3-point function over ${\bf x}_{12}, {\bf x}_{13}$: 
\begin{align} 
\label{GuiLorentz}
    \mathcal{G}(u_i) &= \frac{\kappa_3}{c^{\frac{6-D}{2}}}\, \frac{\mathcal{N}_{\D_1, \D_2, \D_3}}{\prod_{i=1}^3\alpha_D\left(\D_i\right)} \int_{-\infty}^\infty \prod_{i=1}^2 d^{D-2} {\bf x}_{i}\frac{\left|{\bf x}_{12}\right|^{4-D}}{\displaystyle{\prod_{i<j}}\left(-c^2 u_{ij}^2+{\bf x}_{ij}^2+i \varepsilon\right)^{\D_{ij}} }
\end{align}
This is a highly non-trivial integral, but is precisely the kind of integrals encountered in computing two-loop Feynman diagrams and we can use the plethora of well known techniques available for such computations. To start with, let us first check if the scaling works out correctly to give a non-zero and finite Carrollian limit. Note that the change of variables ${\bf x}_1 \to c({\bf x}_1 - {\bf x}_3)$ and ${\bf x}_2 \to c({\bf x}_2 - {\bf x}_3)$ changes the form of the integral to: 
\begin{align}
    \mathcal{G}(u_i) &= \frac{c^{\frac{D-6}{2}}}{\prod_{i=1}^3\alpha_D\left(\D_i\right)} \frac{1}{c^{\Sigma_\D - D}} \int_{-\infty}^\infty\prod_{i=1}^2 d^{D-2} {\bf x}_{i} \frac{\kappa_3\mathcal{N}_{\D_1, \D_2, \D_3} \left|{\bf x}_{12}\right|^{4-D}
 \, }{\left(- u_{12}^2 + \left| {\bf x}_{12}\right|^2 \right)^{\D_{12}} \left(- u_{23}^2 + |{\bf x}_2|^2\right)^{\D_{23}} \left(- u_{13}^2 + |{\bf x}_1|^2\right)^{\D_{13}}}
\end{align}
Recall that each of the $\alpha_D(\D_i)$ in the expression above itself has a scaling of $c^{-\D_i+\frac{D-2}{2}}$, which ensures that $\mathcal{G}(u_i)$ is independent of $c$, and hence is non-zero and finite in the Carrollian limit. To explicitly compute the integral, we will use the Mellin-Barnes integral representation, 
\begin{equation}
    \frac{1}{(A+B)^\nu} = \frac{1}{2\pi i \Gamma(\nu)} \int_{-i\infty}^{i\infty} du\, \Gamma(\nu + u)\Gamma(-u) \frac{B^u}{A^{\nu + u}},
\end{equation}
for two of the three factors in the denominator to get:  
\begin{align}
    \mathcal{G}(u_i) &= \mathcal{K}_3 \prod_{i=1}^2 \int_{-i\infty}^{i\infty}  dv_i \int_{0}^{i\infty} d^{D-2} {\bf x}_{i}   \frac{\Gamma(\D_{13} + v_1)\Gamma(-v_1) \Gamma(\D_{12} + v_2)\Gamma(-v_2) (-u_{13}^2)^{v_1} (-u_{12}^2)^{v_2}}{|{\bf x}_1|^{2\D_{13} + 2v_1} |{\bf x}_{12}|^{2\D_{12} + 2v_2+D-4} \left( -u_{23}^2 + |{\bf x}_2|^2 \right)^{\D_{23}} }
\end{align}
where $\mathcal{K}_3$ is defined to be: 
\begin{equation}\label{kappa3}
    \mathcal{K}_3 = - \frac{c^{\frac{3D-6}{2} - \Sigma_\D}}{\prod_{i=1}^3\alpha_D\left(\D_i\right)} \frac{\kappa_3\mathcal{N}_{\D_1, \D_2, \D_3}}{4\pi^2 \Gamma(\D_{12}) \Gamma(\D_{13})} = \frac{2^{\Sigma_{\D}-D-6}}{\pi^{\frac{D+11}{2}}} \Gamma\left(\frac{\Sigma_{\D}-(D-1)}{2}\right) \Gamma\left(\D_{23}\right),
\end{equation}
and is actually independent of $c$. The integral over ${\bf x}_i$ is well known and $\mathcal{G}(u_i)$ evaluates to : 
\begin{align}
      \mathcal{G}(u_i) &= \frac{\tilde{\mathcal{K}}_3}{\left(-u_{23}^2\right)^{\frac{\Sigma_{\D}-D}{2}}} \prod_{i=1}^2 \int_{-i\infty}^{i\infty}  dv_i  \nonumber \left(\frac{u_{13}}{u_{23}}\right)^{2v_1} \left(\frac{u_{12}}{u_{23}}\right)^{2v_2} \gamma\left(v_1, v_2\right) 
\end{align}
with $\tilde{\mathcal{K}}_3 = \frac{\mathcal{K}_3 \pi^{D-2}}{\Gamma\left(\D_{23}\right)\Gamma\left(\frac{D}{2}-1\right) }$ and 
\begin{align}
\gamma\left(v_1, v_2\right) =&\pi \frac{\Gamma\left(v_1+v_2+\D_1-1\right)\Gamma\left(\frac{D-2}{2}-v_1-\D_{13}\right)\Gamma\left(v_1+v_2+\frac{\Sigma_{\D}}{2} - \frac{D}{2}\right )}{\sin\pi \left(v_2+\D_{12}\right) \Gamma\left(v_2+\D_{12} + \frac{D-4}{2}\right) } \Gamma(-v_1) \Gamma(-v_2).
\end{align}
The evaluation of the $v_1$, $v_2$ integral now requires a careful study of the singularity structure of $\gamma\left(v_1, v_2\right)$ We will deform and close the $v_1$ and $v_2$ contour in the right half-plane. The only possible singularities in the right half-plane come from the presence of $\Gamma$ functions, which have singularities for non-negative integers. In particular, note that for $D \geq 4$, when $\D_{1} > \frac{D}{4}$ and $\Sigma_\Delta > D$, we have no singularities coming from $\Gamma(v_1 + v_2 + \D_1 - 1)$ and $\Gamma\left(v_1 + v_2 + \frac{\Sigma_\Delta}{2} - \frac{D}{2}\right)$. Hence, for simplicitly, we will restrict ourselves to this regime. We can then see that the only possible singularities one can encounter are from the following $\Gamma$ functions: $\Gamma(-v_1)$, $\Gamma(-v_2)$, $\Gamma\left(\frac{D-2}{2}-v_1-\D_{13}\right)$, and $\Gamma(1 - v_2 - \D_{12})$. Now note that the integrand then has singularities on the right-half plane at: 
\begin{equation}
    v_1 \in \mathds{Z}_{\geq 0} \ \ \ \ , \ \ \ \ v_2 \in \mathds{Z}_{\geq 0} \ \ \ \ , \ \ \ \ v_1 + \D_{13} - \frac{D-2}{2} \in \mathds{Z}_{\geq 0} \ \ \ \ , \ \ \ \ v_2 + \D_{12} - 1 \in \mathds{Z}_{\geq 0}
\end{equation}
$\Gamma$ functions have simple pole singularities. Hence, the integrand has simple pole singularities when the singularities coming from the above $\Gamma$ functions are distinct and non-overlapping. This requires the assumption that: 
\begin{equation}
    \D_{13} - \frac{D-2}{2} \notin \mathds{Z} \ \ \ \ , \ \ \ \ \D_{12} \notin \mathds{Z}
\end{equation}
If these constraints are not satisfied then we have overlapping singularities and then one needs to take into account the $\mathcal{O}(1)$ term in he Laurent series expansion of $\Gamma$ functions, in addition to the simple pole term. However, for the purpose of demonstrating an explicit computation, we will suppose that the above holds and that the $\Gamma$ function singularities are all distinct. Noting that $\text{Res} \ \Gamma(z)|_{z = -n} = \frac{(-1)^n}{n!}$, the double integral evaluates to the double sums:  
\begin{align} 
\label{eq:ctildedoublesumrep}
   \mathcal{G}(u_i) =- \frac{ 4\pi^2\tilde{\mathcal{K}}_3}{(-u_{23}^2)^{\frac{\Sigma_\Delta - D}{2}}} \sum_{n_1, n_2 = 0}^\infty A_1^{n_1} A_2^{n_2} &\left[ P_{n_1, n_2} + Q_{n_1, n_2} A_1^{\frac{D-2}{2}-\D_{13}} \right. \\
   &\nonumber \left. \qquad +  R_{n_1, n_2} A_2^{1-\D_{12}} + S_{n_1,n_2}  A_1^{\frac{D-2}{2}-\D_{13}} A_2^{1-\D_{12}}\right],
\end{align}
where $   A_1 = \left( \frac{u_{13}}{u_{23}} \right)^2,  A_2 = \left( \frac{u_{12}}{u_{23}} \right)^2$ and 
\begin{align}
    P_{n_1,n_2} &=\frac{(-1)^{n_1+n_2}}{n_1! n_2!}\frac{\Gamma(\D_{12} + n_2)}{\Gamma\left(\D_{12} + n_2 + \frac{D-4}{2}\right)}\Gamma\left(\frac{D-2}{2}-\D_{13}-n_1\right) \\
    &\nonumber \qquad\qquad \times \Gamma\left(1-\D_{12}-n_2\right) \Gamma(\D_1+n_1+n_2-1)\Gamma\left(\frac{\Sigma_\Delta}{2} + n_1 + n_2 - \frac{D}{2} \right) \\ 
  \nonumber Q_{n_1, n_2} &= \frac{(-1)^{n_1+n_2}}{n_1! n_2!}\frac{\Gamma(\D_{12} + n_2)}{\Gamma\left(\D_{12} + n_2 + \frac{D-4}{2}\right)}\Gamma\left(\D_{13}-n_1-\frac{D-2}{2}\right) \\
  &\nonumber \qquad\qquad \times \Gamma(1-\D_{12}-n_2) \Gamma(\D_2+n_1+n_2-1)\Gamma\left(\D_{12} + n_1 + n_2 + \frac{D-4}{2} \right) \\
  \nonumber R_{n_1,n_2} &=\frac{(-1)^{n_1+n_2}}{n_1! n_2!} \frac{\Gamma(1 + n_2)}{\Gamma\left(n_2 + \frac{D-2}{2}\right)}\Gamma\left(\frac{D-2}{2}-\D_{13}-n_1\right) \\
  & \nonumber \qquad\qquad \times \Gamma(\D_{12}-n_2-1) \Gamma\left(\D_3+n_1+n_2-\frac{D-2}{2}\right)\Gamma\left(\D_{13} + n_1 + n_2\right) \\
  S_{n_1, n_2} &=\frac{(-1)^{n_1+n_2}}{n_1! n_2!}\frac{\Gamma(1 + n_2)}{\Gamma\left(n_2 + \frac{D-2}{2}\right)} \Gamma\left(\D_{13}-n_1-\frac{D-2}{2}\right) \Gamma(\D_{12}-n_2-1) \nonumber\\
  &\nonumber \qquad\qquad \times \Gamma\left(n_1 + n_2 + \frac{D-2}{2}\right)\Gamma\left(\D_{23} + n_1 + n_2 \right) .
\end{align}
To get a closed-form expression for the above sums, we first note that the $n_1$, $n_2$ - dependent $\Gamma$ functions can be written in terms of Pocchammer symbols using:
\begin{equation}
    (q)_k = \frac{\Gamma(q+k)}{\Gamma(q)}
\end{equation}
Secondly, note that when $D = 4$, the structure of the summation resembles the definition of the Appell hypergeometric function $F_4$: 
\begin{equation}
    F_4(a,b;c,d;A_1, A_2) = \sum_{n_1 = 0}^\infty \sum_{n_2 = 0}^\infty \frac{(a)_{n_1 + n_2} (b)_{n_1 + n_2}}{(c)_{n_1} (d)_{n_2}} \frac{A_1^{n_1} A_2^{n_2}}{n_1! n_2!}
\end{equation}
For general $D$, it looks slightly more complicated due to the presence of $|\textbf{x}_{12}|^{D-4}$ in the ansatz \eqref{eq:3ptposspace}. However, there is a well-known function that generalizes the generalized hypergeometric function, known as the Kampe de Feriet function \cite{https://doi.org/10.1002/nme.1620140114}: 
\begin{equation}
    F^{p;r,k}_{q;s,l}(\Vec{\textbf{a}_{(p)}}, \Vec{\textbf{b}_{(q)}}; \Vec{\textbf{c}_{(r)}}, \Vec{\textbf{d}_{(s)}}; \Vec{\textbf{c}'_{(k)}}, \Vec{\textbf{d}'_{(l)}}; A_1, A_2) = \sum_{n_1 = 0}^\infty \sum_{n_2 = 0}^\infty \frac{(\Vec{\textbf{a}_{(p)}})_{n_1+n_2} (\Vec{\textbf{c}_{(r)}})_{n_1} (\Vec{\textbf{c}'_{(k)}})_{n_2}}{(\Vec{\textbf{b}_{(q)}})_{n_1+n_2} (\Vec{\textbf{d}_{(s)}})_{n_1} (\Vec{\textbf{d}'_{(l)}})_{n_2}}  \frac{A_1^{n_1} A_2^{n_2}}{n_1! n_2!}
\end{equation}
where $\Vec{\textbf{v}}_{(i)} = \{v_1, v_2, \ldots v_i\}$. $(\Vec{\textbf{v}_{(i)}})_k$ is a vector with $i$ components and $ (\Vec{\textbf{v}_{(i)}})_{k} $ denotes the product of the Pocchammer symbols $(v_1)_k (v_2)_k \ldots (v_i)_k$. After some simplification, we express $\mathcal{G}\left(u_i\right)$ in terms of the Kampe de Feriet function $F^{2;0,1}_{0;1,2}$ as: 
\begin{align} 
\label{finalLorentz}
    \mathcal{G}(u_i) &= -\frac{4\pi^2\tilde{\mathcal{K}}_3}{(-u_{23}^2)^{\frac{\Sigma_\Delta - D}{2}} }\left[C_1  F_1 + C_2 \left(\frac{u_{13}}{u_{23}}\right)^{D-2-2\D_{13}} F_2  +  C_3 \left(\frac{u_{12}}{u_{23}}\right)^{2-2\D_{12}}  F_3  \right. \\
    \nonumber & \left. \hspace{6cm}  + C_4 \left(\frac{u_{13}}{u_{23}}\right)^{D-2-2\D_{13}} \left(\frac{u_{12}}{u_{23}}\right)^{2-2\D_{12}}  F_4\right] ,
\end{align}
where 
\begin{align}
\label{eq:ctildeCi}
    &C_1 = \frac{\Gamma(\D_{12})}{\Gamma\left(\D_{12} + \frac{D-4}{2}\right)}\Gamma\left(\frac{D-2}{2}-\D_{13}\right) \Gamma(1-\D_{12}) \Gamma(\D_1-1)\Gamma\left(\frac{\Sigma_\Delta - D}{2} \right), \\
    &C_2 =\frac{\Gamma(\D_{12})}{\Gamma\left(\D_{12} + \frac{D-4}{2}\right)} \Gamma\left(\D_{13}-\frac{D-2}{2}\right) \Gamma(1-\D_{12}) \Gamma(\D_2 - 1)\Gamma\left(\D_{12} + \frac{D-4}{2}\right), \nonumber \\
    &C_3= \frac{1}{\Gamma \left(\frac{D-2}{2}\right)} \Gamma\left(\frac{D-2}{2}-\D_{13}\right) \Gamma(\D_{12}-1) \Gamma\left(\D_3-\frac{D-2}{2}\right)\Gamma\left(\D_{13}\right), \nonumber \\
    &C_4 =  \Gamma\left(\D_{13}-\frac{D-2}{2}\right) \Gamma(\D_{12}-1) \Gamma(\D_{23}) ,
\end{align}
and
\begin{align} 
\label{eq:ctildeF}
   & F_1 = F^{2;0,1}_{0;1,2} \left( \left\lbrace  \D_1 - 1, \frac{\Sigma_\Delta - D}{2}\right\rbrace,\lbrace\textbf{0}\rbrace; \lbrace\textbf{0}\rbrace,  \left\lbrace \D_{13}-\frac{D-4}{2}\right\rbrace; \left\lbrace \textbf{0}\right\rbrace, \left\lbrace\D_{12} + \frac{D-4}{2}, \textbf{0}\right\rbrace ; \left(\frac{u_{13}}{u_{23}}\right)^2, \left(\frac{u_{12}}{u_{23}}\right)^2 \right)  \nonumber \\
    &F_2 = F^{2;0,1}_{0;1,2} \left(\left\lbrace \D_2 - 1, \D_{12} + \frac{D-4}{2}\right\rbrace ,\lbrace\textbf{0}\rbrace; \lbrace\textbf{0}\rbrace, \left\lbrace \frac{D}{2} - \D_{13}\right\rbrace ; \left\lbrace\textbf{0}\right\rbrace, \left\lbrace \D_{12} + \frac{D-4}{2}, \textbf{0}\right\rbrace ; \left(\frac{u_{13}}{u_{23}}\right)^2, \left(\frac{u_{12}}{u_{23}}\right)^2 \right) \nonumber\\
    &F_3 = F^{2;0,1}_{0;1,2} \left(\left\lbrace \D_3 -\frac{D-2}{2}, \D_{13}\right\rbrace , \lbrace\textbf{0}\rbrace; \lbrace\textbf{0}\rbrace, \left\lbrace \D_{13}-\frac{D-4}{2}\right\rbrace ; \left\lbrace 1\right\rbrace, \left\lbrace 2-\D_{12}, \frac{D-2}{2}\right\rbrace; \left(\frac{u_{13}}{u_{23}}\right)^2, \left(\frac{u_{12}}{u_{23}}\right)^2 \right) \nonumber \\ 
    &F_4 =  F^{2;0,1}_{0;1,2} \left(\left\lbrace \frac{D-2}{2}, \D_{23}\right\rbrace , \lbrace\textbf{0}\rbrace; \lbrace\textbf{0}\rbrace, \left\lbrace\frac{D}{2}-\D_{13}\right\rbrace; \left\lbrace 1\right\rbrace ,\left\lbrace 2-\D_{12}, \frac{D-2}{2}\right\rbrace ; \left(\frac{u_{13}}{u_{23}}\right)^2,  \left(\frac{u_{12}}{u_{23}}\right)^2 \right) 
\end{align}
As mentioned earlier, when $D = 4$, this reduces to a sum of the Appell $F_4$ hypergeometric functions. \\

\bibliographystyle{style}
\bibliography{Biblio}

\end{document}